\documentclass[]{interact}

\usepackage{epstopdf}% To incorporate .eps illustrations using PDFLaTeX, etc.
\usepackage{subfigure}% Support for small, `sub' figures and tables

\usepackage{natbib}% Citation support using natbib.sty
\bibpunct[, ]{(}{)}{,}{a}{}{,}% Citation support using natbib.sty
\renewcommand\bibfont{\fontsize{10}{12}\selectfont}% Bibliography support using natbib.sty

\usepackage{textcomp}
\usepackage[noend]{algpseudocode}
\usepackage{algorithmicx,algorithm}
\usepackage{threeparttable}
\usepackage{multirow}
\usepackage{type1cm}
\theoremstyle{plain}% Theorem-like structures

\theoremstyle{definition}

\theoremstyle{remark}

\begin{document}

\articletype{ARTICLE MANUSCRIPT}

\title{Extending regionalization algorithms to explore spatial process heterogeneity}

\author{
\name{Hao Guo\textsuperscript{a,b}, Andre Python\textsuperscript{c} and Yu Liu\textsuperscript{a,b}\thanks{CONTACT Yu Liu. Email: liuyu@urban.pku.edu.cn}}
\affil{\textsuperscript{a}Institute of Remote Sensing and Geographical Information Systems, School of Earth and Space Sciences, Peking University, Beijing, China; \textsuperscript{b}Beijing Key Lab of Spatial Information Integration and Its Applications, Peking University, Beijing, China;\textsuperscript{c}Center for Data Science, Zhejiang University, Hangzhou, China}
}

\maketitle

%\linenumbers

\begin{abstract}

In spatial regression models, spatial heterogeneity may be considered with either continuous or discrete specifications. The latter is related to delineation of spatially connected regions with homogeneous relationships between variables (spatial regimes). 
Although various regionalization algorithms have been proposed and studied in the field of spatial analytics, methods to optimize spatial regimes have been largely unexplored. 
In this paper, we propose two new algorithms for spatial regime delineation, two-stage K-Models and Regional-K-Models. We also extend the classic Automatic Zoning Procedure to spatial regression context. 
The proposed algorithms are applied to a series of synthetic datasets and two real-world datasets. Results indicate that all three algorithms achieve superior or comparable performance to existing approaches, while the two-stage K-Models algorithm largely outperforms existing approaches on model fitting, region reconstruction, and coefficient estimation. 
Our work enriches the spatial analytics toolbox to explore spatial heterogeneous processes.

\end{abstract}


\section{Introduction}

Along with spatial dependency, spatial heterogeneity is one of the two fundamental properties of spatial data, commonly observed in both natural and social phenomena \citep{Ans88,Sh05,FoSa22}. Spatial heterogeneity can be observed in varying attribute values across space, and in the geographic processes that generate the attribute data. As a result, principles and laws in social and environmental sciences usually do not hold across large spatial domains and scales, in contrast to physics or chemistry \citep{GoLi21}. Replicability, which is the ability to obtain consistent results using similar data and methods, is therefore difficult to reach strictly in geographic studies \citep{SuKe21}. Yet replicability may be partially achieved by allowing some aspects of a geographical model (e.g. its estimated parameters) to vary across space, while keeping its structure (e.g. the set of features) replicable. This partial replicability refers to the concept of ``weak replicability'' \citep{GoLi21,LGL22}. 

A global model fitted for a whole study area may estimate an overall trend across space, thus unable to capture spatial heterogeneity. As a result, the modeling accuracy (measured by errors between observed data and model output) may be inadequate. In contrast, models with a continuous heterogeneity specification (also known as local models) may explicitly account for spatial heterogeneity by allowing model parameters to vary with geographic locations, which is in line with the concept of weak replicability \citep{FoBr99}. Geographically weighted regression (GWR, \cite{BFC96,FYK17}) is a typical approach in this category. Yet the flexibility gained by introducing local parameters ineluctably affects the ability of local models to capture general patterns. The pointwise model representation is not parsimonious, as it entails a substantially larger number of parameters than a global model. 

Models with a discrete heterogeneity specification, referred to as spatial regime models \citep{Ans10}, offer an intermediate approach compromising between accuracy and parsimony. In this framework, the study area is divided into a set of zones, which are usually required to be spatially contiguous, and a set of model parameters is calibrated for each zone. The spatial contiguity constraint reflects the principle of spatial dependency formulated in Tobler's first law of geography \citep{Tob70}, which asserts that near locations tend to exhibit similar geographical processes (relationship between variables). Although the region delineation may be specified \emph{a priori}, it can be more appealing to detect regions with homogeneous processes from data, especially when the purpose is to reveal spatial variation of relationships rather than examine differences between \emph{pre-existing} regions.

Nevertheless, the analytical task of region delineation is complicated by its large solution space \citep{Kea75}. Regionalization, or spatial clustering methods, are designed to derive homogeneous spatially connected regions by grouping neighboring spatial units with similar attributes. The $p$-regions formulations based on mixed integer programming (MIP) provide an approach to solve this problem exactly. Yet it is computationally expensive and only applicable to datasets with very limited size \citep{DCM11}. Therefore, heuristic methods have become the mainstream of research, represented by the automatic zoning procedure (AZP) \citep{Open77, OpRa95}, spatial `k’luster analysis by tree edge removal (SKATER) \citep{ANC06}, and regionalization with dynamically constrained agglomerative clustering and partitioning (REDCAP) \citep{Guo08}. 

Note that delineation of spatial regimes is an intrinsically different task from regionalization. The latter considers heterogeneity of \emph{form} (attribute values), while the former considers heterogeneity of \emph{process} (relationships between variables). A heterogeneous process usually produces heterogeneous attributes, yet heterogeneous attributes are not necessarily generated by a heterogeneous process. As opposed to regionalization, few studies have addressed the optimization problem to detect spatially connected regions with homogeneous processes. This task is crucial to environmental and social studies, as process heterogeneity between regions is studied in various fields such as oceanography \citep{LSM21} and epidemiology \citep{XLS11}, and may contribute to a better understanding of the earth surface system. An exception is the recent extension of the SKATER algorithm to spatial regression \citep{AnAm21, VPB22}. Yet the solution space is limited by the minimum spanning tree (MST), which may lead to suboptimal results.

In this paper, we further explore the extension of regionalization algorithms to spatial regime delineation. Based on the spatially implicit K-Means and the spatially explicit Regional-K-Means algorithms, we propose two new algorithms, namely \textit{two-stage K-Models} and \textit{Regional-K-Models}. We also extend the classic automatic zoning procedure (AZP) to spatial regression context. We compare the performance of the three algorithms with two existing approaches, GWR-Skater \citep{HBHL13} and Skater-reg \citep{AnAm21}, in model fitting, region reconstruction, and coefficient estimation. Results on synthetic datasets indicate that the performance of all three introduced algorithms is comparable to that of GWR-Skater and Skater-reg, while the two-stage K-Models largely outperforms the baseline approaches. The proposed algorithms are applied to two real-world datasets, including the Georgia census dataset and the King County house price dataset, illustrating their usefulness to capture spatial heterogeneous processes.

\section{Related works}

\subsection{Regionalization algorithms}

Regionalization can be viewed as a specific spatial clustering procedure aiming at aggregating spatial units into geographically connected regions \citep{Wei22}. After the regionalization procedure, each unit is assigned to a unique region, and each region contains at least one unit \citep{DRS07}. The number of regions is often predefined, although some framework such as max-p-regions \citep{DAR12} treats it as an endogenous variable. The criteria used to identify the regions may include equity or threshold of attributes\footnote{For example, the population in each region is required to be as similar as possible or above a predefined value (see \cite{DAR12,FoSp14,WRK21}). } \citep{DAR12}, compactness of geometry \citep{LCG14}, spatial auto-correlation \citep{OpRa95}, or goodness-of-fit of a global model\footnote{Note that the optimization of spatial regimes differs from \cite{Open78}, where spatial units are aggregated into areas, and each area is treated as an observation in a global regression model.} \citep{Open78}.

A basic criterion for regionalization is homogeneity, which assumes that spatial units in a region are similar (in a set of attributes), while those in different regions are distinct. This is in line with generic clustering analysis in unsupervised machine learning. Considering a set of spatial areal units $u_i (i=1,2,\dots,n)$, assume $\mathbf{x}_i$ is the attribute vector associated with $u_i$. Two formulations of the objective function are commonly used in current methods. The first is the sum of within-region variability across all delineated regions: 
\begin{equation}
    \label{eq:rego1}
    \mathcal{L(\mathcal{R})}= \sum_{j=1}^p \sum_{1 \le i_1 < i_2 \le n} I[u_{i_1},u_{i_2}\in R_j]\|\mathbf{x}_{i_1}-\mathbf{x}_{i_2}\|^2,
\end{equation}
where $\mathcal{R}=\{R_j\}_{j=1}^p$ is a region partition of $\{u_i\}_{i=1}^n$; $I[\text{cond}]$ is an indicator function, which takes 1 if the condition (cond) is true, 0 otherwise; $\|\cdot\|$ represents a vector norm, commonly formulated as Euclidean $L_2$-norm\footnote{Note that in Equation \ref{eq:rego1}, the number of considered unit pairs in the sum is $\sum_{j=1}^M \binom{|R_j|}{2}$, which is smaller if $|R_j| (j=1,\dots,M)$ are close to each other. Hence the objective function might favor solutions whose regions have similar numbers of units.}. The second is the sum of squared deviations (SSD) from the cluster center (mean of the attribute vectors), in line with clustering algorithms such as K-Means:
\begin{equation}
    \label{eq:rego2}
    \mathcal{L(\mathcal{R})}= \sum_{j=1}^p \sum_{i=1}^n I[u_{i}\in R_j]\; \|\mathbf{x}_{i}-\boldsymbol{\mu}_{j}\|^2
\end{equation}
where
\begin{equation}
    \boldsymbol{\mu}_{j}=\frac{1}{|R_j|}\sum_{i=1}^n I[u_i\in R_j] \; \mathbf{x}_i
\end{equation}
Alternatively, the objective may be formulated from an information-theoretic perspective, where homogeneity is measured with Bregman information \citep{Cho17}, or description length \citep{Kir22}. Based on how spatial contiguity is treated, existing algorithms may be classified into two main categories, spatially implicit methods and spatially explicit methods \citep{DRS07}.  

Spatially implicit methods first apply a generic clustering algorithm to the set of attribute vectors, without necessarily defining geographically connected regions. To impose spatial contiguity, non-connected regions are then broken into connected parts in the post-processing stage \citep{OpWy95}. Spatially implicit methods are generally less preferred in literature \citep{AJA21}, as they do not allow strict control of the number of regions \citep{DRS07}. Moreover, such approaches tend to produce undesired small regions, as clusters in the attribute space could be segmented into pieces in geographic space \citep{Guo08}. Manual refinement may be performed to mitigate both issues, yet it cannot be exempt from subjectivity. 

Spatially explicit methods consider spatial contiguity during the regionalization process, which ensures the connectivity of regions without requiring post-processing operations. \cite{DCM11} formulated the regionalization problem as three mixed integer programming (MIP) models. Using commercial optimization software, the optimal solutions to these formulations are guaranteed to be found. However, such exact methods are computationally expensive and therefore applicable to small datasets only. While heuristic methods do not guarantee finding the optimal solution, their computational advantages make them more suitable for large datasets. \cite{Open77} proposed the AZP algorithm, which uses an iterative process to improve an initial solution by moving a unit from one region to another while ensuring spatial contiguity. This algorithm was improved by incorporating intelligent optimization techniques, such as simulated annealing and tabu search \citep{OpRa95}. \cite{ANC06} proposed SKATER algorithm which operates by removing edges from the minimum spanning tree. \cite{Guo08} proposed REDCAP algorithm, which combines spatially constrained agglomerative clustering and tree edge removal. The latter two algorithms may produce more accurate and stable results compared to AZP \citep{AJA21}. \cite{AlGe06} proposed a multidirectional optimum ecotope-based algorithm (AMOEBA) for spatial hotspot detection on univariate data, which has been also applied to exhaustive region partition \citep{WGH10}. As a component of the Python package spopt \citep{spopt}, Rey developed Regional-K-Means algorithm, which follows a workflow similar to K-Means but incorporates a spatial contiguity check before each move. 

After a regionalization process, the statistical significance of heterogeneity between regions (spatially stratified heterogeneity) could be assessed with the $q$-statistic and related test introduced by \cite{WZF16}. 

\subsection{Spatial regime delineation}

Regime regression allows regression coefficients to vary across groups of observations. Each regime is defined as a non-empty subset of observations, and each observation belongs to exactly one regime. A spatial regime consists of spatially contiguous observations \citep{AnAm21}. A regime linear regression model may have its slopes, intercept, or both, varying across regimes. The Chow test \citep{Chow60} is used to assess whether there is a significant difference in regression coefficients between regimes. Regimes have been incorporated in models with spatially lagged variables \citep{ElFr09, AnRe14}, and a spatialized Chow test has been proposed by \cite{Ans90}. 

Regimes can be defined with exogenous information (e.g. administrative districts), or based on the similarity of variables (e.g. clustering analysis). For example, \cite{ElFr09} used a threshold of governors' support rate to divide areas into two regimes. \cite{ELB06} used per capita GDP (one of the independent variables), with the aid of Moran scatterplot. However, the derived regimes may not necessarily reflect heterogeneity of the spatial process, as similarity of associations (relationship between variables) are not considered in these approaches.

For consistency with the meaning of regimes, we argue that spatial regime delineation should consider homogeneity of the underlying spatial process that generates the data, which can be assessed by evaluating robustness of a regression or probability model across a region. In this sense, spatial regime delineation is an intrinsically different task from regionalization. For example, consider a spatial cross-sectional dataset with a dependent variable $y$ and an independent variable $x$, which take values $y_i$ and $x_i$ at spatial unit $u_i$, respectively. When considering the regime delineation for a simple linear model, we consider a region $R$ as homogeneous if there exists a constant $\beta$ such that $y_i = \beta x_i$ holds for all units $u_i$ within $R$, even if the values of $x$ and $y$ may exhibit significant spatial variation across the region. Hence, heterogeneity of attribute values are not necessarily generated by heterogeneity of the process. 

Two categories of methods have been developed to derive regimes with homogeneous processes. The first category of methods is based on local regression models such as GWR. \cite{BCMM16} used spatial functional regression to estimate local coefficients, which are subsequently grouped with Ward’s clustering algorithm. \cite{ABP17} and \cite{BBP17} developed iterated spatially weighted regression (ISWR), which transforms continuous GWR parameter surfaces into discrete regimes by adjusting the local weights. Yet both approaches do not strictly apply the spatial contiguity constraint. To ensure region contiguity, regionalization algorithms could be applied on local coefficients instead of generic clustering. For example, \cite{HBHL13} took a GWR-Skater approach, applying SKATER algorithm to local regression coefficients derived with GWR. The second type obtains the coefficients and the regime allocation jointly. \cite{AnAm21} and \cite{VPB22} independently proposed an extension of SKATER algorithm to delineate spatial regimes, which is named Skater regression (Skater-reg) or SkaterF. Based on a minimum spanning tree (MST), the algorithm removes tree edges iteratively to minimize total regression error while satisfying region contiguity. However, as distances in attribute space are used as edge weights, the MST is still based on similarity of variables (attribute homogeneity). This may lead to logical inconsistency with the objective to minimize regression error (process homogeneity). Such an MST may restrict the solution space, leading to sub-optimal results. \cite{LiSa19} proposed the spatially clustered coefficients regression (SCC), which adopts $L_1$ regularization to identify regions with homogeneous coefficients. However, the solution space is still restricted by a predefined minimum spanning tree based on Euclidean distance, and the method may produce more regions than expected \citep{LSM21}. 

Bayesian spatial partition methods differ from the frequentist (also known as classical) statistical approaches in that they explicitly represent the uncertainty of regression coefficients and region delineation through a probabilistic model for the data generation process. The solution space for region delineation is modeled with Voronoi polygons \citep{KHR00,DeHo01}, or random spanning trees \citep{TAL19,LSM21}. The latter is preferred for its ability to represent latent regions of arbitrary shapes. Methods that use random spanning trees also overcome the limitation induced by the single spanning tree considered in SKATER and SCC. However, the state-of-the-art methods in this direction still face challenges in terms of computation cost \citep{LSM21}. 

\section{Methodology}

Consider $n$ spatial areal units $\{u_i\}_{i=1}^n$ in a geographically connected study area\footnote{Throughout the paper, we describe the case of lattice data (spatial data on areal units). Our approach is also applicable to point observation data after building adjacency (with k-nearest neighbors (KNN) or Delaunay triangulation, for example).}, with a set of $m$ independent variables $\mathbf{x}_i =(x_{i1},\dots,x_{im})^\mathrm{T}$ and a dependent variable $y_i$ for each unit $u_i$. We consider rook contiguity, so two units are neighbors if their boundaries share a common edge. Given the number of regions $p$, we aim to partition the units into $p$ geographically connected regions $\mathcal{R}=\{R_j\}_{j=1}^p$, each with a homogeneous relationship between the independent variables $\mathbf{x}$, and the dependent variable $y$. We assume that all regions use the same class of model, while the parameters are allowed to vary across regions. Let $\boldsymbol{\theta}_j$ denote the estimated parameters in region $R_j$, and $\Theta=\{\boldsymbol{\theta}_1.\dots, \boldsymbol{\theta}_p\}$. The region scheme is optimized by minimizing the total sum of squared residuals (SSR):
\begin{equation}
    \min_{\mathcal{R},\Theta} \sum_{j=1}^{p}\sum_{i=1}^{n} I[u_i \in R_j] \; (y_i-f(\mathbf{x}_i,\boldsymbol{\theta}_j))^2
\end{equation}
where $f(\mathbf{x},\boldsymbol{\theta})$ is the predicted value of the regression model $f$ with independent variables $\mathbf{x}$ and parameters $\boldsymbol{\theta}$. In other words, we jointly estimate the region-specific regression parameters, along with the delineation of regions. Although our method does not restrict the form of model $f$, we consider the basic case of multiple linear regression throughout this paper. Hence $f(\mathbf{x})=\alpha+\mathbf{x}^\mathrm{T}\boldsymbol{\beta}$, with parameters composed of an intercept $\alpha$ and a coefficient vector $\boldsymbol{\beta}$ that may vary across regions. Hence the region-specific parameters $\boldsymbol{\theta}_j=(\alpha_j, \boldsymbol{\beta}_j)$. 

The proposed algorithms to optimize spatial regimes are built upon a model calibration procedure, which is used to estimate and update region-specific parameters:
\begin{equation}
\label{eq:localfit}
    \boldsymbol{\theta}_j = \mathop{\arg\min}\limits_{\boldsymbol{\theta}} \sum_{i=1}^{n} I[u_i \in R_j] \; (y_i-f(\mathbf{x}_i,\boldsymbol{\theta}))^2
\end{equation}
For multiple linear regression models, this corresponds to the ordinary least squares (OLS) procedure, which is used throughout this paper.

We investigate three algorithms to optimize spatial regimes, including two spatially explicit algorithms, AZP and Regional-K-Models, and one spatially implicit algorithm, two-stage K-Models. All considered algorithms start with an initial feasible solution and execute an iterative process. To improve the solution, units are moved from one region to another in each iteration, and related regression coefficients are updated accordingly.  In K-Models and Regional-K-Models, a unit can only be moved into the region with its ``closest'' regression model (i.e. the model with the lowest residual). However, in AZP, a move is allowed as long as it decreases the SSR.

A minimum size constraint should be applied to each region for validity of the model parameter estimation. For multiple linear regressions, a region should have at least $m+1$ observations to ensure the uniqueness of OLS estimation. We introduce a parameter $\text{min}\_\text{obs}$ in all the algorithms, which stands for the minimum number of units in each region. 

To generate the initial solution, each region is initialized with a different random seed picked from $\{u_i\}_{i=1}^n$. Next, for each region, if it has unassigned neighboring units, one of them is randomly picked and assigned to the region. This step is repeated until every unit is assigned to a region. The procedure ensures the spatial contiguity of each region. If the initial solution does not satisfy the minimum size constraint, the procedure will restart with another set of random seeds\footnote{This usually happens when  $\text{min}\_\text{obs}$ is close to $n/p$, where $p$ is the number of regions. Given $\text{min}\_\text{obs} \ll n/p$, this issue does not create problems, as observed in our experiments.}.

\subsection{Two-stage K-Models algorithm}\label{sec:kmodels}

We propose two-stage K-Models algorithm as an extension of the classical K-Means clustering algorithm \citep{MQ67}. The idea is similar with spatially clustered regression (SCR) proposed by \cite{SuMu21}. The difference lies in the specification of the spatial contiguity constraint. SCR introduced a penalty term, encouraging neighboring units to be in the same cluster. Such soft constraint does not ensure region contiguity. Inspired by two-stage clustering algorithms such as Chameleon \citep{KHK99}, our two-stage method first partition units into ``micro-clusters'', then merge them into geographically connected regions. Hence, region contiguity is guaranteed. 

The K-Models algorithm starts with a randomly generated initial region scheme with $K$ regions $\{R_1,\dots,R_K\}$, where $K>p$ is a hyperparameter standing for the number of micro-clusters. For each initial region $R_j$, a set of model parameters $\boldsymbol{\theta}_j$ is initialized. Note that the choice of $K$ does not influence the number of produced regions. The performance of the algorithm is not sensitive to the choice of $K$ in a reasonable interval. From our experiments, setting $2p \le K\le 4p$ seems to work well. If $K$ is too small, the algorithm may fail to produce the required number of regions, while a too large $K$ may have minor negative effects on the performance. 

In the first stage (partition stage), the initial solution is improved by reallocating the spatial units iteratively. Each iteration includes two steps. First, for each unit, the regression residual is calculated with each set of coefficients $\{\boldsymbol{\theta}_1,\dots, \boldsymbol{\theta}_K\}$, and the unit is moved to the region whose model fits it best (unless the move would break the minimum size constraint). Second, the coefficients of each region are updated. The algorithm stops if no unit is moved during an iteration, or when the maximum number of iterations is reached. Note that the total SSR is non-increasing in both steps, yet the resulting regions after the first stage are not necessarily connected. The pseudo-code for the first stage is provided in Algorithm \ref{KModels}. The time complexity of each iteration is $O(Kmn+m^2(m+n))$.

\begin{algorithm}[t]
\caption{Two-stage K-Models: partition stage} 
\hspace*{0.02in} {\bf Input:} 
Dataset $\{(\mathbf{x}_i,y_i)\}_{i=1}^n$; adjacency matrix of the units $\{u_i\}_{i=1}^n$; number of clusters $K$; minimum cluster size $\text{min}\_\text{obs}$; maximum iterations $L$.\\
\hspace*{0.02in} {\bf Output:} 
Regions $\{R_j\}_{j=1}^K$ and regression parameters $\{\boldsymbol{\theta}_j\}_{j=1}^K$.
\begin{algorithmic}[1]
\State Generate initial regions $\{R_j\}_{j=1}^K$ and estimate initial parameters $\{\boldsymbol{\theta}_j\}_{j=1}^K$;
\State stable = False;
\State $t = 0$;
\While{not stable and $t<L$} 
\State $R_j'=R_j$, $j=1,\dots,K$; 
\For{$i=1,2,\dots,n$} 
    \State Find $d_i \in \{1,\dots,K\}$ such that $u_i \in R_{d_i}$;
    \If{$|R_{d_i}|>\text{min}\_\text{obs}$}
    \State $r_i =\operatorname{\arg\min}\limits_{1\le j \le K} |y_i-f(\mathbf{x}_i, \boldsymbol{\theta}_j)|$; 
    \State $R_{d_i}= {R}_{d_i}\backslash \{u_i\}$, $R_{r_i}= {R}_{r_i}\cup \{u_i\}$.
    \EndIf
\EndFor
\For{$j=1,2,\dots,K$}
    \State $\boldsymbol{\theta}_j=\mathop{\arg\min}\limits_{\boldsymbol{\theta}}\sum\limits_{i=1}^n I[u_i\in R_{j}] (y_i-f(\mathbf{x}_i, \boldsymbol{\theta}_j))^2$;
\EndFor
\If{$R_j=R_j'$, with $j=1,2,\dots,K$}
    \State stable = True
\EndIf
\State $t=t+1$;
\EndWhile
\State \Return $\{R_j,\boldsymbol{\theta}_j\}_{j=1}^K$.
\end{algorithmic}
\label{KModels}
\end{algorithm}

The second stage (merge stage) ensures that the required number of regions $p$ is satisfied, as well as the contiguity of each region. First, if a region is disconnected, it is divided into connected parts, each of which is considered as a new region. Then regions with the number of units fewer than $\text{min}\_\text{obs}$ are merged with neighboring regions so that the minimum size constraint is satisfied\footnote{If a region with inadequate units has two or more neighboring regions, we select the neighbor which minimizes the total SSR after the merge.}. A new set of regression coefficients is estimated for each region. Normally, the number of regions remains larger than $p$ after this step\footnote{When $\text{min}\_\text{obs}$ is too large or $K$ is too small (close to $p$), exceptions may occur that the number of regions is less than $p$, hence the algorithm cannot produce the required number of regions by merging ``micro-clusters''. This issue can be solved by adjusting $\text{min}\_\text{obs}$ and $K$. }. Finally, we examine each pair of neighboring regions, and calculate the decrease in the SSR if they are merged (a new set of model parameters is fitted for the merged region). The pair of regions that produces the largest decrease are merged into one region. This step is repeated until the required number of regions is reached. 

\subsection{AZP algorithm}

Unlike most regionalization algorithms, AZP offers more flexibility on the objective function, which makes it adaptable to various zoning criteria. We apply AZP to optimize spatial regimes. Starting from an initial solution which satisfies the spatial contiguity constraint, AZP tries to improve the solution by moving a region's neighboring unit into the region. The AZP procedure is described in Algorithm \ref{azp}.

\begin{algorithm}[t]
\caption{AZP} 
\hspace*{0.02in} {\bf Input:} 
Dataset $\{(\mathbf{x}_i,y_i)\}_{i=1}^n$; adjacency matrix of the units $\{u_i\}_{i=1}^n$; number of regions $p$; minimum region size $\text{min}\_\text{obs}$; maximum iterations $L$.\\
\hspace*{0.02in} {\bf Output:} 
Regions $\{R_j\}_{j=1}^p$ and estimated parameters $\{\boldsymbol{\theta}_j\}_{j=1}^p$.
\begin{algorithmic}[1]
\State Generate initial regions $\{R_j\}_{j=1}^p$ and estimate initial parameters $\{\boldsymbol{\theta}_j\}_{j=1}^p$;
\State stable = False;
\State $t=0$;
\While{not stable and $t<L$}
\State stable = True; 
\For{$j=1,2,\cdots,p$}
    \State $P=\varnothing$;
    \For{$v$ in the set of units which are contiguous with $R_j$}
        \State Find the region $R_d$ such that $v \in R_d$;
        \If {$|R_d|>\text{min}\_\text{obs}$, $R_d\backslash \{v\}$ is connected, and moving $v$ to $R_j$ decreases SSR}
        \State $P=P\cup \{v\}$;
        \EndIf
    \EndFor
    \If{$P\neq \varnothing$}
    \State stable = False;
    \State Randomly choose $v \in P$;
    \State Move $v$ into $R_j$, and update the regression parameters.
    \EndIf
\EndFor
\State $t=t+1$;
\EndWhile
\State \Return $\{R_j,\boldsymbol{\theta}_j\}_{j=1}^p$.
\end{algorithmic}
\label{azp}
\end{algorithm}

\begin{figure}
\centering
\subfigure[]{\resizebox*{6cm}{!}{\includegraphics{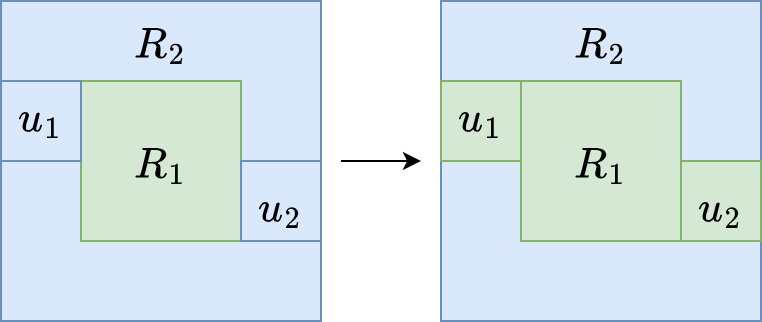}}}
\hspace{2em}
\subfigure[]{\resizebox*{6cm}{!}{\includegraphics{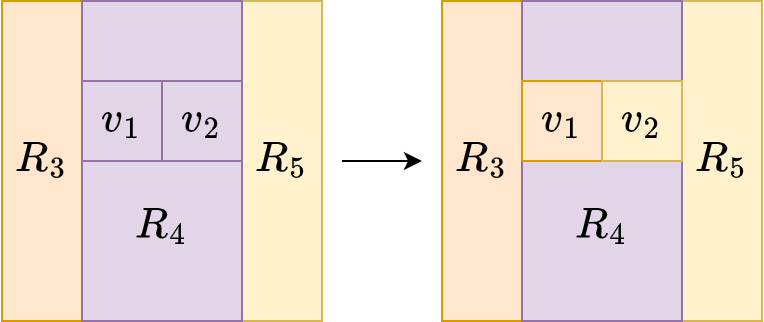}}}
\caption{Examples of spatial contiguity breaks that may occur if two moves are performed simultaneously. (a) $R_1$, $R_2$ are regions, and units $u_1$, $u_2$ initially belong to $R_2$. In AZP algorithm, moving $u_1$ or $u_2$ into $R_1$ satisfies the contiguity constraint, while moving both of them into $R_1$ simultaneously makes $R_2$ disconnected. (b) In Regional-K-Models, a similar situation may occur for region $R_4$, if $v_1$ is moved into region $R_3$, and simultaneously, $v_2$ is moved into $R_5$, although either move is in principle valid. } 
\label{cont}
\end{figure}

Three conditions need to be met to allow moving a unit from one region to another: (a) the donor region still satisfies the minimum size constraint after the move; (b) the donor region remains connected after giving out the unit; (c) the move leads to a decrease in regression error. While checking the first condition is straightforward, the second condition involves the verification of graph connectivity. A breadth-first search approach has a time complexity of $O(|E|)$, where $E$ is the set of edges in the adjacent graph. Assessing the validity of the third condition for a spatial unit $v$ in the neighborhood of $R_j$ requires fitting regression models for $R_j\cup \{v\}$ and $R_d\backslash \{v\}$. The time complexity is $O(m(n+m))$\footnote{Let $n_r$ denote the number of units in the region. The OLS estimation of the coefficient vector is $\boldsymbol{\beta}= (\mathbf{X}^{\mathrm{T}}\mathbf{X})^{-1}\mathbf{X}^{\mathrm{T}}\mathbf{y}$, where $\mathbf{X}$ is the $n_r\times (m+1)$ matrix of independent variables, $\mathbf{y}$ is the $n_r$-dimensional vector of dependent variable. Here the intercept is included in $\boldsymbol{\beta}$ by adding an independent variable with constant value 1. By applying the Sherman-Morrison formula \citep{Bar51} to update the $(\mathbf{X}^{\mathrm{T}}\mathbf{X})^{-1}$ term, the time complexity can be reduced from $O(m^2(n_r+m))$ to $O(m(n_r+m))$.}, which is comparable with the region contiguity condition assuming $m\ll n$ and $|E|=O(n)$ for the adjacency graph. Hence, the minimum size constraint should be verified first, followed by the other two conditions which are more time-consuming.  In the presence of two or more valid neighboring units of $R_j$, only one unit is moved, as it may affect the validity of subsequent moves. Figure \ref{cont}a illustrates the potential issues associated with simultaneous moves. The time complexity of each iteration is roughly bounded by $O(pn(\vert E\vert+m(m+n))$.

\subsection{Regional-K-Models algorithm}

We propose to extend Regional-K-Means algorithm \citep{spopt}, which is further referred to as Regional-K-Models. In contrast to the two-stage K-Models algorithm, spatial contiguity is checked before each move, which makes Regional-K-Models spatially explicit. The procedure of Regional-K-Models is described in Algorithm \ref{rkm}. Similar with the partition stage of two-stage K-Models, the objective function is non-increasing in both unit moving and model updating. 

\begin{algorithm}[t]
\caption{Regional-K-Models} 
\hspace*{0.02in} {\bf Input:} 
Dataset $\{(\mathbf{x}_i,y_i)\}_{i=1}^n$; adjacency matrix of the units $\{u_i\}_{i=1}^n$; number of regions $p$; minimum region size $\text{min}\_\text{obs}$; maximum iterations $L$.\\
\hspace*{0.02in} {\bf Output:} 
Regions $\{R_j\}_{j=1}^p$ and estimated parameters $\{\boldsymbol{\theta}_j\}_{j=1}^p$.
\begin{algorithmic}[1]
\State Generate initial regions $\{R_j\}_{j=1}^p$ and estimate initial parameters $\{\boldsymbol{\theta}_j\}_{j=1}^p$;
\State stable = False;
\State $t=0$;
\While{not stable and $t<L$}
\State stable = True;
\State $P=\varnothing$;
\For{$i=1,2,\dots,n$}
    \State Find $d_i \in \{1,\dots,K\}$ such that $u_i \in R_{d_i}$;
    \If{$|R_{d_i}|>\text{min}\_\text{obs}$}
    \State $r_i =\operatorname{\arg\min}\limits_{1\le j \le K} |y_i-f(\mathbf{x}_i, \boldsymbol{\theta}_j)|$; 
    \If {$d_i \neq r_i$ and $R_{d_i}\backslash \{u_i\}$ is connected}
        \State $P=P\cup \{u_i\}$;
    \EndIf
    \EndIf
    
\EndFor
\If{$P\neq \varnothing$}
    \State stable = False;
    \State Randomly choose $u_i \in P$;
    \State Move $u_i$ into $R_{r_i}$, and update the related model parameters.
\EndIf
\State $t=t+1$;
\EndWhile
\State \Return $\{\mathcal{R}_j,\theta_j\}_{j=1}^K$.
\end{algorithmic}
\label{rkm}
\end{algorithm}

As a unit is allowed to be moved into its neighboring region with the best model fit only, the range of potential candidate moves is rather restricted compared with AZP. Similar to AZP, only one candidate move is performed at a time, in order to prevent a potential break of the contiguity constraint induced by simultaneous moves (see illustrated example in Figure \ref{cont}b). The time complexity of each iteration is bounded by $O(pmn+n\vert E\vert+m^2(m+n))$.

\section{Experiments on synthetic data}\label{sec:synthetic}

\subsection{Simulation design}\label{sec:simset}

To assess the performance of the proposed algorithms, we performed a set of regression experiments on synthetic data. The datasets are generated on a $25\times25$ regular grid, where each grid cell represents a spatial unit. The relationship between $\mathbf{x}$ and $y$ exhibits strict stratified heterogeneity. That is, the true regression coefficients are identical in each predefined region and distinct in different regions. This enables us to test whether the proposed algorithms can reconstruct the underlying homogeneous regions associated with geographical processes, given the spatial distributions of $\mathbf{x}$ and $y$. 

\begin{figure}
\centering
\resizebox*{12cm}{!}{\includegraphics{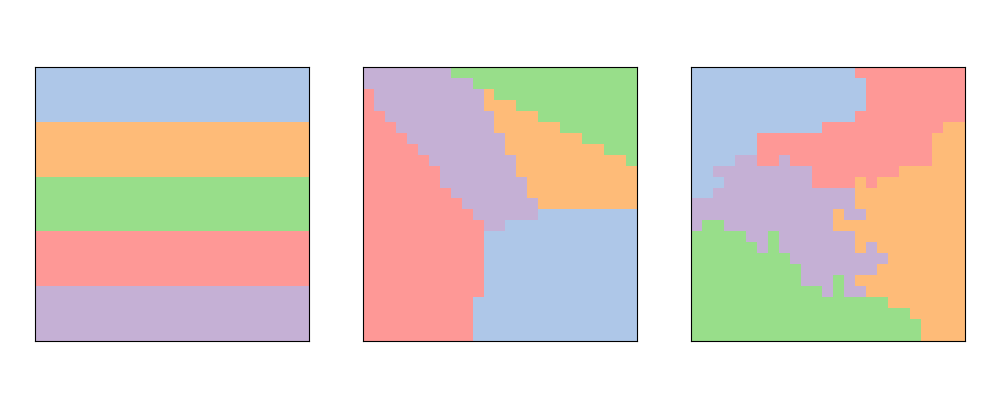}}
\caption{Examples of true region schemes on a 25$\times$25 grid from the three generation approaches: \textit{Rectangular}, \textit{Voronoi}, and \textit{Arbitrary} (from left to right). Each region is assigned a different set of $(\beta_1, \beta_2)$.} 
\label{figdis}
\end{figure}

For each cell, each independent variable $x_i$ is generated from an independent uniform distribution, and the dependent variable $y$ is generated from a linear model, with predefined linear coefficients. The data generating process can be expressed as follows:
\begin{align}
    &y = \beta_0 + \beta_1 x_1 + \beta_2 x_2 + \epsilon\\
    &x_1, x_2\sim \mathcal{U}[0,1), \, i.i.d.\\
    &\epsilon \sim \mathcal{N}(0,\sigma^2), \, i.i.d.
\end{align}
where $\mathcal{U}[0,1)$ is the uniform distribution on $[0,1)$; $\mathcal{N}(0,\sigma^2)$ is the normal distribution with mean 0 and variance $\sigma^2$. The grid is partitioned into five connected regions $\{R_1,\dots,R_5\}$. Each of the covariate coefficients $\beta_{1}, \beta_{2}$ take different values from $\{-2, -1, 0, 1, 2\}$ in the five regions\footnote{Let $\beta_{i,j}$ denote the value of coefficient $\beta_i$ in region $R_j$. In each simulation, the list $(-2, -1, 0, 1, 2)$ is randomly shuffled twice, and used as $(\beta_{1,1},\dots,\beta_{1,5})$ and $(\beta_{2,1},\dots,\beta_{2,5})$, respectively.}. We set $\beta_0 = 0$ for all regions. To assess the capability of the proposed algorithms to detect regions in different shapes, three approaches are used to generate latent region schemes: 
\begin{enumerate}
    \item \textit{Rectangular}: The grid is partitioned into five rectangles, each with $5\times 25$ units;
    \item \textit{Voronoi}: Five cells are randomly picked as seeds, and their corresponding Voronoi polygons produce five spatially connected regions; 
    \item \textit{Arbitrary}: Five cells are randomly picked as seeds. The seeds randomly grow into five spatially connected regions. 
\end{enumerate}
Furthermore, we enforce a constraint ensuring that each region contains at least 10 units. Examples from these approaches are shown in Figure \ref{figdis}. 

For each of the three region generation approaches, we generated 50 simulations with the data generating process defined above. Each simulation includes a true region scheme, true coefficients $(\beta_1,\beta_2)$ in each region, and the $\mathbf{x},y$ values for each unit. For the \textit{Rectangular} dataset, the region scheme is constant across all simulations, while each simulation has different region schemes in \textit{Voronoi} and \textit{Arbitrary} datasets. We set $\sigma=0.1$. The performance of the proposed algorithms under a higher degree of random noise is investigated in Appendix. 

Two existing approaches are used as baselines, GWR-Skater\footnote{\cite{HBHL13} also applied principal component analysis to the GWR coefficients. This step is skipped, as dimension reduction is not needed in our experiment.} \citep{HBHL13} and Skater-reg \citep{AnAm21}. For all algorithms, we set the number of regions $p=5$, $\text{min}\_\text{obs}=10$\footnote{Different values of $\text{min}\_\text{obs}$ may be used in the two stages of K-Models algorithm. Here $\text{min}\_\text{obs}=10$ is used for the merge stage, while $\text{min}\_\text{obs}$ in the partition stage is the number of independent variables plus 1 throughout this paper.}. For two-stage K-Models, we set $K=20$. For GWR-Skater, we apply an adaptive bisquare kernel with corrected Akaike information criterion (AICc) used as the criterion to select the bandwidth \citep{FYK17}. The data generating process and three proposed algorithms are implemented in Python 3.9. We use the linear regression function provided by scikit-learn \citep{sklearn}, the GWR implementation in mgwr \citep{OLK19}, the SKATER implementation in spopt, and the Skater-reg implementation in spreg. The mgwr, spopt, spreg packages are modules of PySAL, a family of Python packages for spatial data science \citep{RAA22}. All experiments on synthetic data are performed on a computer with dual Intel Xeon Gold 5118 CPUs (2.30GHz) and 256GB of memory. 

\begin{figure}
\centering
\resizebox*{9cm}{!}{\includegraphics{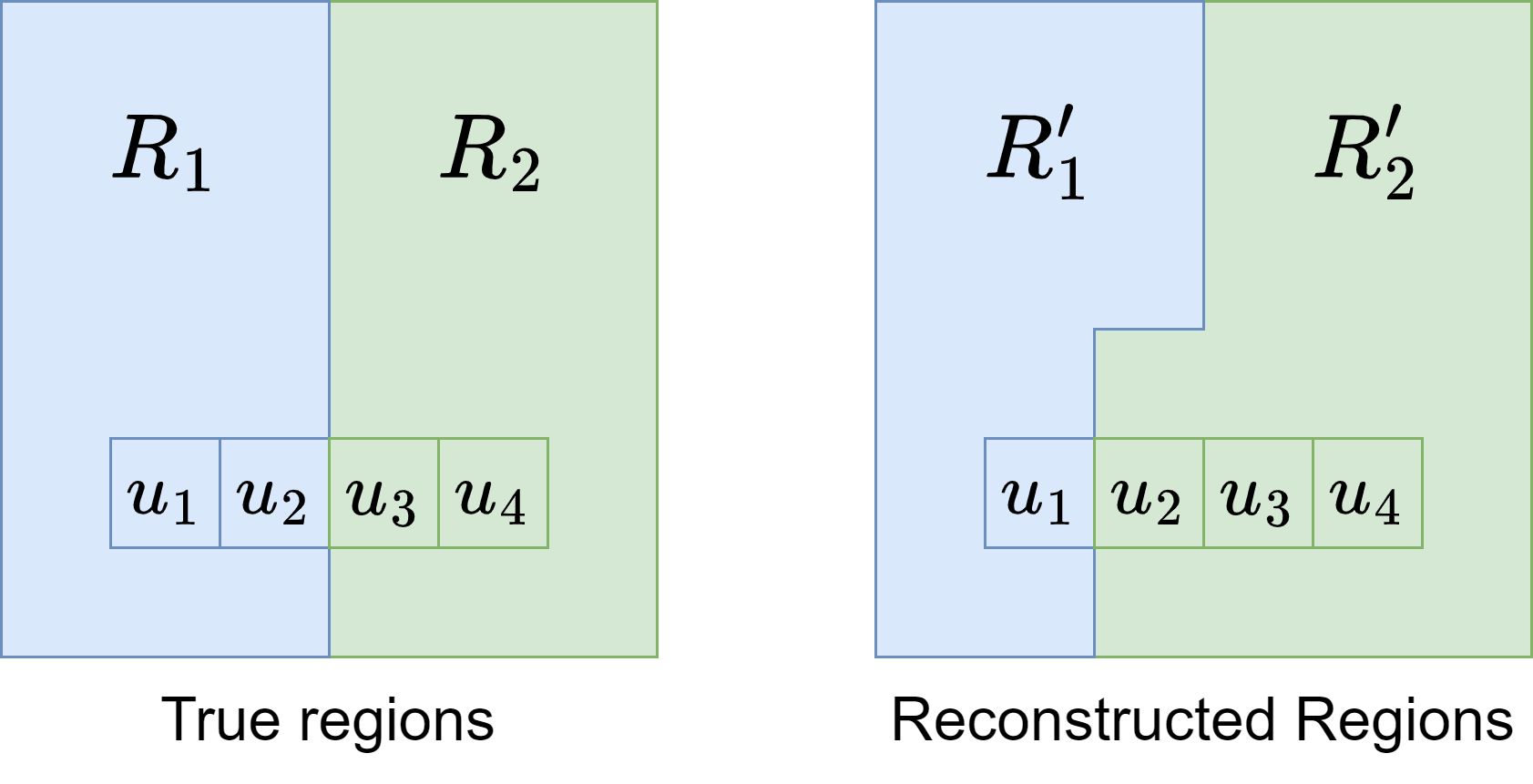}}
\caption{Four possible situations for a pair of spatial units in the calculation of RI. $R_1$, $R_2$ are true regions, $R_1'$, $R_2'$ are reconstructed regions, $u_i(i=1,2,3,4)$ are units. $(u_3,u_4)$ is a true positive (TP) pair, $(u_1,u_2)$ is a false negative (FN) pair, $(u_2,u_3)$,$(u_2,u_4)$ are two false positive (FP) pairs, $(u_1,u_3)$,$(u_1,u_4)$ are two true negative (TN) pairs.} 
\label{fig:randex}
\end{figure}

We adopt six metrics, reflecting three aspects of algorithm performance:
\begin{itemize}
    \item Model fitting: the total sum of squared residuals (SSR); 
    \item Region reconstruction: Rand index (RI) and normalized mutual information (NMI); 
    \item Coefficient estimation: mean absolute error (MAE) of $\beta_0$, $\beta_1$, and $\beta_2$. The absolute coefficient error of each spatial unit is averaged. 
\end{itemize}
RI and NMI are used to quantify the discrepancy between the true regions and reconstructed regions. Four possible situations may arise for each pair of spatial units in a region delineation result (Figure \ref{fig:randex}):
\begin{itemize}
\item True positive (TP): units are in the same region in both the true and reconstructed schemes.
\item False negative (FN): units are in the same region in the true scheme, but mistakenly grouped into different regions by the algorithm. 
\item False positive (FP):  units are in different regions in the true scheme, but mistakenly grouped into the same region by the algorithm. 
\item True negative (TN): units are in different regions in both the true and reconstructed schemes.
\end{itemize}
RI \citep{Ran71} calculates the proportion of correctly grouped pairs of units as follows:
\begin{equation}
    \text{RI} = \frac{TP+TN}{TP+FN+FP+TN}.
    \label{eq:rand}
\end{equation}
RI can take values between 0 and 1. Higher values indicate a better region reconstruction. RI takes the value 1 if all reconstructed regions are identical to the true regions. NMI measures the discrepancy between two region schemes from an information theory perspective \citep{VEB10}. Assume $\mathcal{R}=\{R_j\}_{j=1}^r$ are the true underlying regions, and $\mathcal{R}'=\{R_j'\}_{j=1}^p$ are reconstructed regions. The entropy $H(\mathcal{R})$ is defined as follows:
\begin{equation}
    H(\mathcal{R})=-\sum_{j=1}^r \frac{|R_j|}{n}\log \frac{|R_j|}{n},
    \label{eq:ent}
\end{equation}
where $n$ is the total number of units. The mutual information $I(\mathcal{R},\mathcal{R}')$ of $\mathcal{R}$ and $\mathcal{R}'$ is defined as follows:
\begin{equation}
    I(\mathcal{R},\mathcal{R}')=\sum_{j=1}^r\sum_{k=1}^p \frac{|R_j \cap R_k'|}{n}\log \frac{n |R_j \cap R_k'|}{|R_j||R_k'|}.
    \label{eq:mi}
\end{equation}
Based on the concepts above, NMI can be expressed as follows:
\begin{equation}
    \text{NMI}(\mathcal{R},\mathcal{R}') = \frac{I(\mathcal{R},\mathcal{R}')}{H(\mathcal{R})H(\mathcal{R'})}.
    \label{eq:nmi}
\end{equation}
The range of NMI is [0,1]. High NMI values indicate that one region scheme provides a large amount of information for the other, which suggests similarity between the two region schemes.

\subsection{Main Results}\label{sec:mainres}
We report the regime optimization performance of five considered algorithms in Table \ref{tabsim}. The metric values are averaged over 50 simulations in each setting. For all three datasets, two-stage K-Models shows the best overall performance according to all six metrics, largely outperforming other methods in model fitting, region reconstruction, and coefficient estimation. AZP achieves the second lowest SSR in all datasets, yet its performance in region reconstruction and coefficient estimation is inferior to GWR-Skater in the \textit{Rectangular} and \textit{Voronoi} datasets. This may indicate that AZP is more prone to overfitting than GWR-Skater. The performance of Regional-K-Models and Skater-reg is not satisfying, compared to the other algorithms.

\begin{table}
\tbl{Comparison of region delineation performance between algorithms on three synthetic datasets.}
{\begin{tabular*}{\hsize}{@{}@{\extracolsep{\fill}}llrrrrrr@{}} 
\toprule
Dataset & Algorithm & SSR & RI & NMI & MAE\,$\beta_0$ & MAE\,$\beta_1$ & MAE\,$\beta_2$\\ \midrule
\multirow{5}{*}{\textit{Rectangular}} & K-Models & \textbf{21.18} & \textbf{0.9719} & \textbf{0.9061} & \textbf{0.0326} & \textbf{0.1113} & \textbf{0.1157} \\
& AZP & 185.74 	&	0.8316 	&	0.5644 	&	0.1361 	&	0.6578 	&	0.6870 \\
& Reg-K-Models & 319.66 	&	0.7962 	&	0.4733 	&	0.1277 	&	0.8173 	&	0.8398 \\
& GWR-Skater & 215.22	&	0.8728 	&	0.7108 	&	0.1054 	&	0.4537 	&	0.4877 \\
& Skater-reg & 389.89	&	0.6895 	&	0.4428 	&	0.1068 	&	0.8737 	&	0.9282 \\
\midrule
\multirow{5}{*}{\textit{Voronoi}} & K-Models & \textbf{22.24} 	&	\textbf{0.9731} & \textbf{0.9023} &	\textbf{0.0385}	& \textbf{0.1135} 	&\textbf{0.1053} \\
& AZP & 112.54 	&	0.8632 	&	0.6814 	&	0.0869 	&	0.4465 	&	0.4284 \\
& Reg-K-Models & 215.25 	&	0.8452 	&	0.6169 	&	0.1048 	&	0.5541 	&	0.5560 \\
& GWR-Skater & 213.25 	&	0.8830 	&	0.6882 	&	0.1078 	&	0.4105 	&	0.4192 \\
& Skater-reg & 283.52 	&	0.7435 	&	0.5195 	&	0.1011 	&	0.7069 	&	0.6743 \\
\midrule
\multirow{5}{*}{\textit{Arbitrary}} & K-Models & \textbf{42.99} &	\textbf{0.9443} & \textbf{0.8359} &	\textbf{0.0472} & \textbf{0.2014} &	\textbf{0.2003}\\
& AZP & 131.72 	&	0.8638 	&	0.6438 	&	0.0969 	&	0.5191 	&	0.5178 \\
& Reg-K-Models & 271.42 	&	0.8255 	&	0.5456 	&	0.1124 	&	0.6935 	&	0.6954  \\
& GWR-Skater & 260.54 	&	0.8390 	&	0.6060 	&	0.1111 	&	0.5401 	&	0.5277 \\
& Skater-reg & 333.62 	&	0.7564 	&	0.4934 	&	0.1099 	&	0.7723 	&	0.7496 \\
\bottomrule
\end{tabular*}}
\label{tabsim}
\footnotesize
Note: K-Models is short for two-stage K-Models; Reg-K-Models is short for Regional-K-Models. The best metric values in each simulation setting are put in bold.
\end{table}

\begin{figure}
\centering
\resizebox*{12cm}{!}{\includegraphics{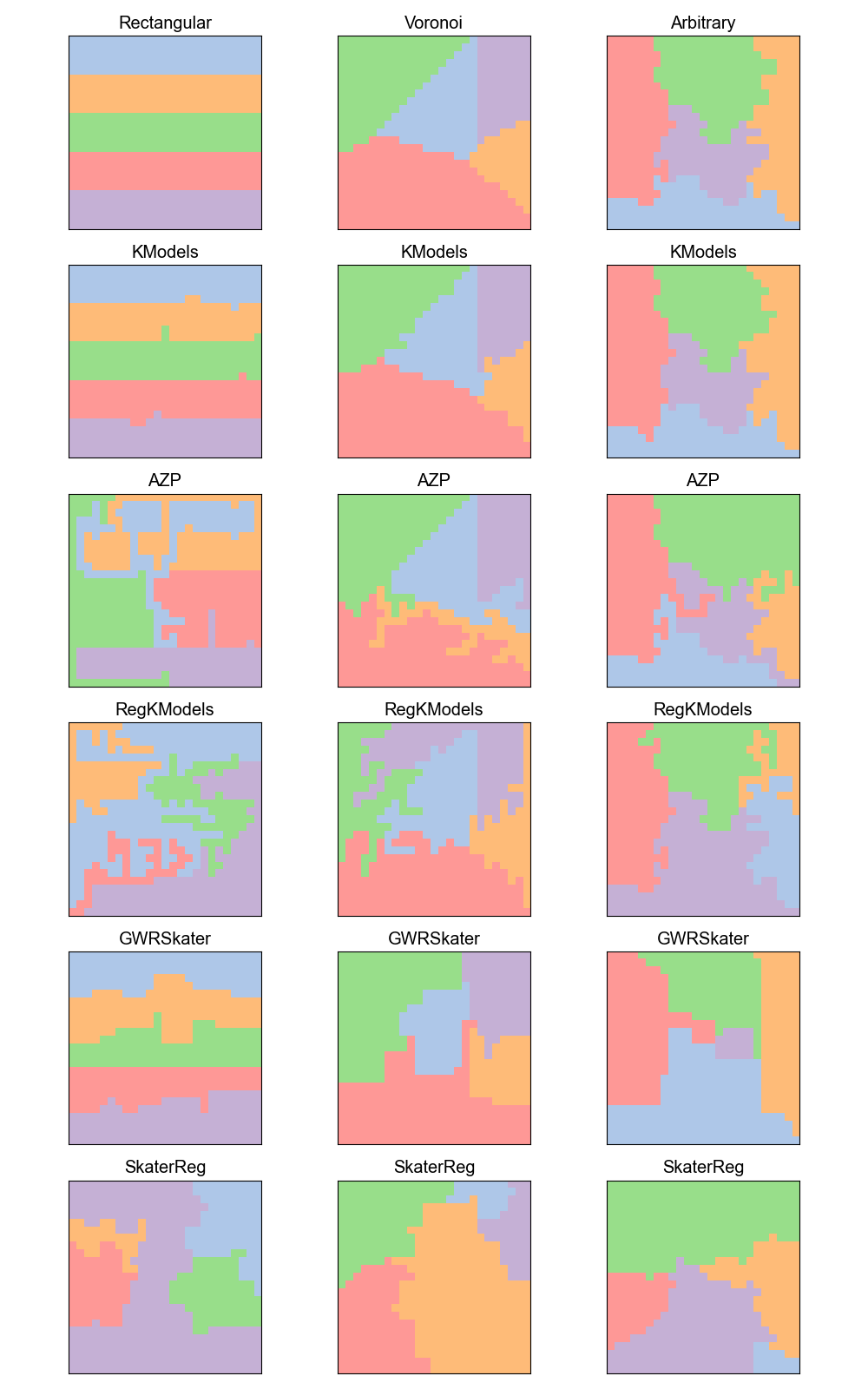}}
\caption{Region delineation results of one illustrative simulation from each dataset. The first row shows the true latent regions associated with regression coefficients. Other rows show the reconstructed regions from five algorithms (two-stage K-Models, AZP, Regional-K-Models, GWR-Skater, and Skater-reg). } 
\label{figsim1}
\end{figure}

Figure \ref{figsim1} illustrates the region delineation results of one illustrative simulation from each of the three datasets. Generally, regions produced by two-stage K-Models are similar with the true regions, except for minor disparities in region boundaries. While GWR-Skater is able to capture the general pattern of regions, it falls short in accurately identifying region boundaries, unlike two-stage K-Models. This may be due to the smoothing of coefficient surfaces induced by GWR. AZP successfully discovers some region boundaries, yet it tends to produce regions with more branches, which do not exist in the true regions. Failure to eliminate such branches may be due to the spatial contiguity constraint, as moving one unit on the branch to another region would break region contiguity. Such conditions also appear in results from Regional-K-Models. Skater-reg often fails to reconstruct the region pattern, which may be caused by a restriction of the solution space induced by a single spanning tree. 

The computational cost of the considered algorithms is reported in Table \ref{tabtime}. The running time is averaged over 50 simulations for each dataset. Despite the improvement in performance, the computational cost of two-stage K-Models is comparable to GWR-Skater and Skater-reg, delineating spatial regimes for a $25\times 25$ grid in several seconds. The running time of Regional-K-Models is about half a minute. The most time-consuming algorithm is AZP, which is also found to be computationally expensive for regionalization \citep{Guo08}. 

\begin{table}
\tbl{Average running time (in seconds) on three synthetic datasets.}
{\begin{tabular*}{\hsize}{@{}@{\extracolsep{\fill}}lrrr@{}} 
\toprule
Algorithm & \textit{Rectangular} & \textit{Voronoi} & \textit{Arbitrary}\\ \midrule
K-Models & 6.03 	&	6.10 	&	6.06  \\
AZP & 82.58 	&	80.50 	&	66.53 \\
Reg-K-Models & 28.55 	&	27.79 	&	24.86  \\
GWR-Skater & 4.82 	&	4.64 	&	4.61  \\
Skater-reg & 4.76 	&	4.66 	&	4.65 \\
\bottomrule
\end{tabular*}}
\label{tabtime}
\footnotesize
Note: K-Models is short for two-stage K-Models; Reg-K-Models is short for Regional-K-Models. 
\end{table}

In the Appendix, we describe supplementary experiments on the algorithm stability, the choice of parameter $K$ in two-stage K-Models, and the effect of a higher degree of random noise. 

\section{Empirical examples}

\subsection{The Georgia dataset}

The Georgia dataset is a sample dataset from the Python library mgwr \citep{OLK19} and has been used in several spatial analytics studies (e.g., \cite{FBC02,Gri08,YFL20}). The data are collected from a population census in 1990 and include socio-economic attributes of 159 counties in Georgia, USA \citep{OLK19}. We consider the regression of the percentage of people with a bachelor’s degree or higher (PctBach), with intercept and explanatory variables including the percentage of people born in a foreign country (PctFB), the percentage of African American (PctBlack), and the percentage of rural residents (PctRural), as in \cite{YFL20}.

As in Section \ref{sec:synthetic}, we compare the three proposed algorithms with GWR-Skater and Skater-reg. Since the number of true regions is unknown, we run the algorithms multiple times, with $p$ varying from 2 to 10. $\text{min}\_\text{obs}=5$ is used for all algorithms. For two-stage K-Models, we set $K=2p$. The three proposed algorithms (two-stage K-Models, AZP, and Regional-K-Models) are repeated 10 times at each $p$, and the solution with the lowest SSR is selected. Repeating runs are unnecessary for GWR-Skater and Skater-reg, as their results do not change on the same data. 

\begin{figure}
\centering
\resizebox*{8cm}{!}{\includegraphics{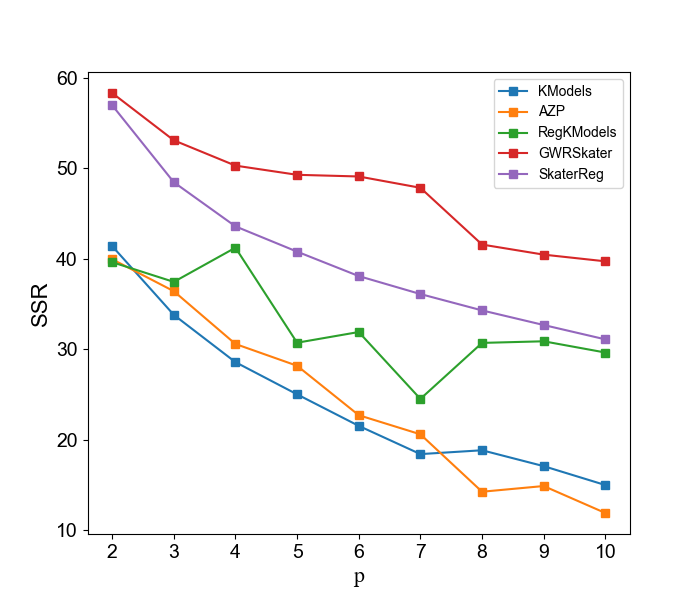}}
\caption{The total sum of squared residuals (SSR) from five considered algorithms at different numbers of regions $p$. For two-stage K-Models, AZP and Regional-K-Models, the lowest SSR in 10 repeating runs is reported. SSR is calculated with standardized independent and dependent variables. } \label{fig:georgp}
\end{figure}

The total SSR of solutions from five considered algorithms at different $p$ is shown in Figure \ref{fig:georgp}. Generally, SSR decreases as the number of regions increases. This reflects the trade-off between accuracy (minimizing modeling errors) and simplicity (minimizing number of parameters) when modeling discrete heterogeneity. The three proposed algorithms consistently outperform GWR-Skater and Skater-reg on this dataset, at different numbers of regions\footnote{Even considering the average SSR rather than the lowest, two-stage K-Models and AZP consistently outperform GWR-Skater and Skater-reg; Regional-K-Models is comparable to Skater-reg and superior to GWR-Skater.}. Unlike results on synthetic data, the performance of AZP is close to two-stage K-Models, and Skater-reg consistently outperforms GWR-Skater. We infer that the relationship between variables in the Georgia dataset does not exhibit strict stratified heterogeneity as specified in synthetic data, which leads to the difference observed in the performance of the algorithms.

\begin{figure}
\centering
\subfigure[Two-stage K-Models]{\resizebox*{4cm}{!}{\includegraphics{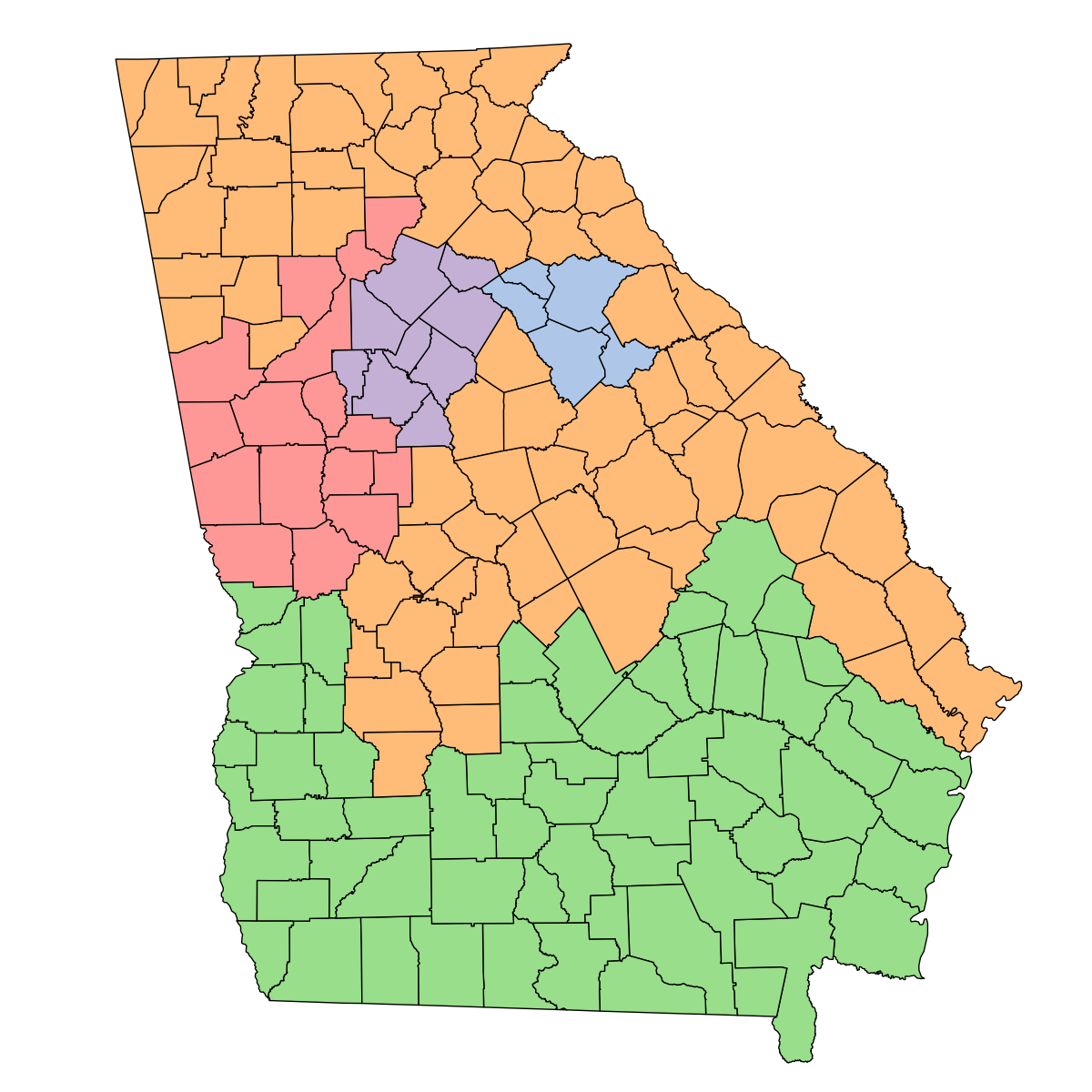}}}
\subfigure[AZP]{\resizebox*{4cm}{!}{\includegraphics{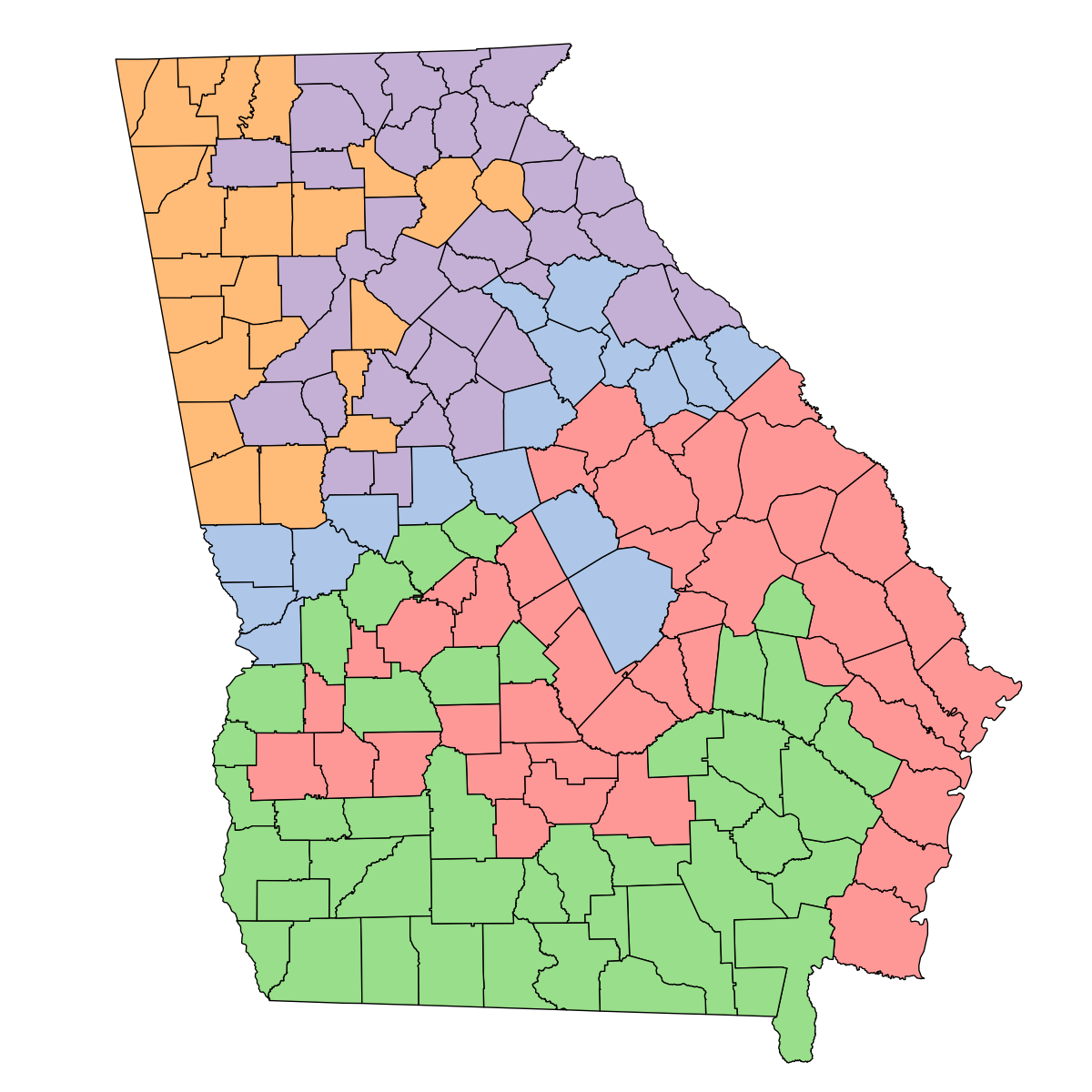}}}
\subfigure[Regional-K-Models]{\resizebox*{4cm}{!}{\includegraphics{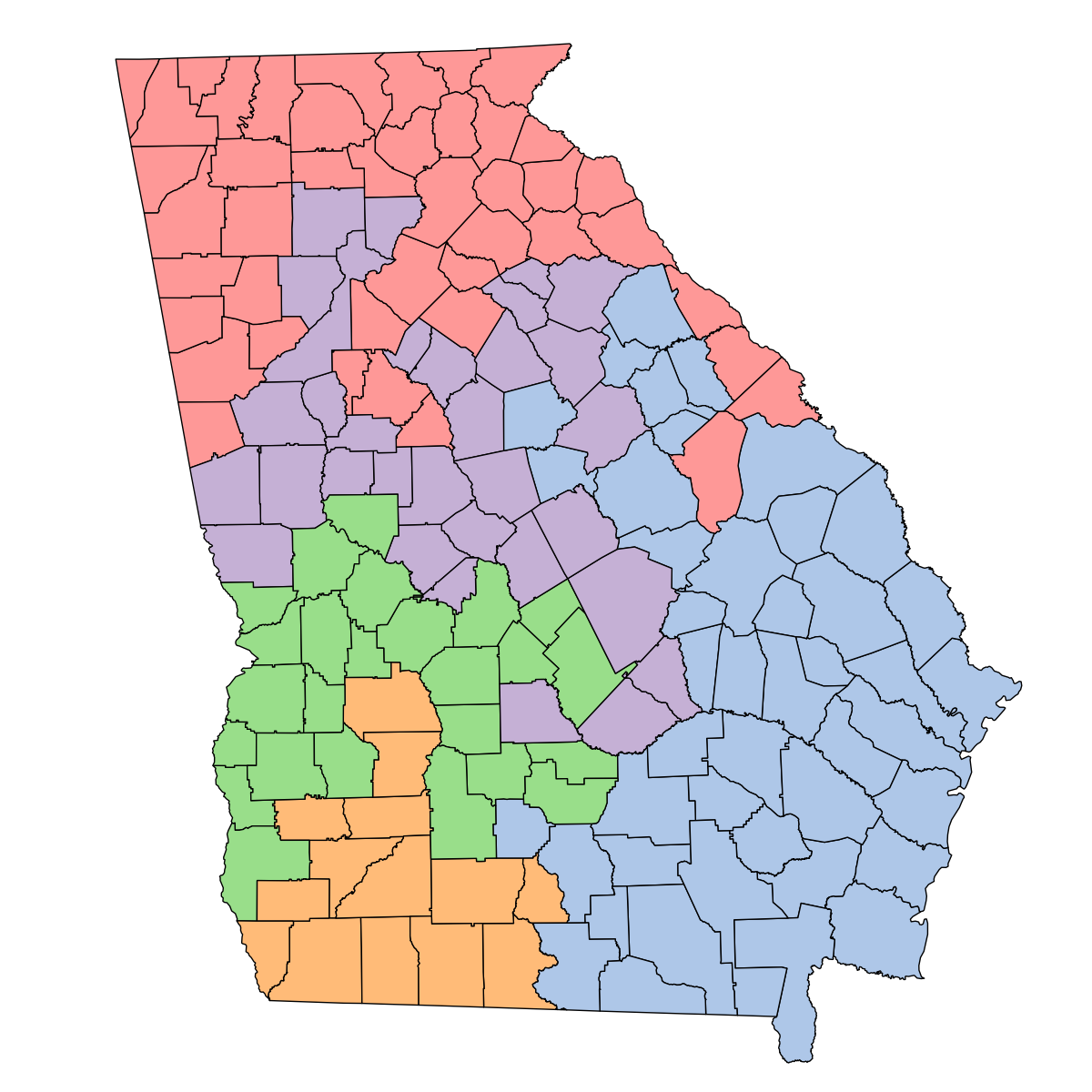}}}
\subfigure[GWR-Skater]{\resizebox*{4cm}{!}{\includegraphics{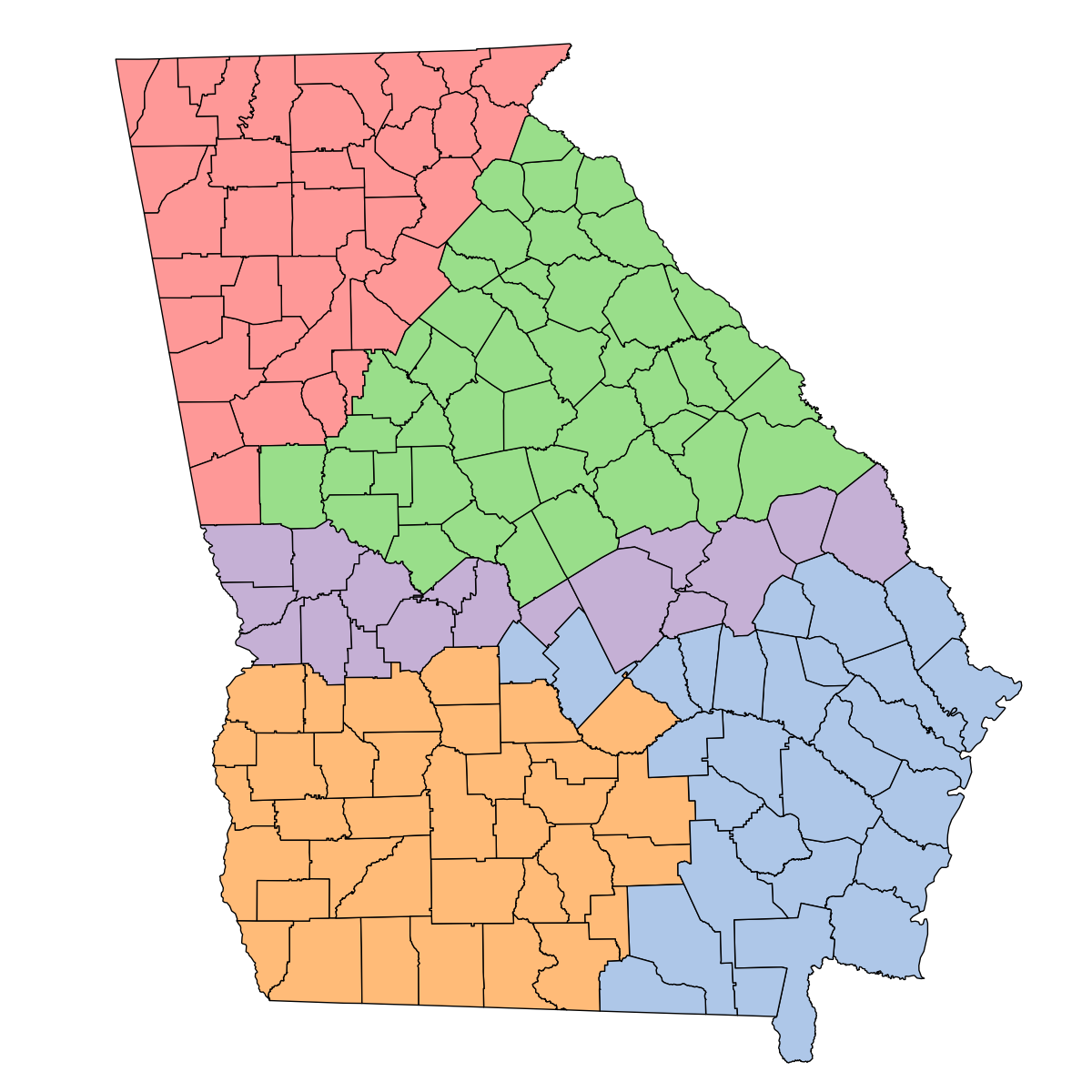}}}
\subfigure[Skater-reg]{\resizebox*{4cm}{!}{\includegraphics{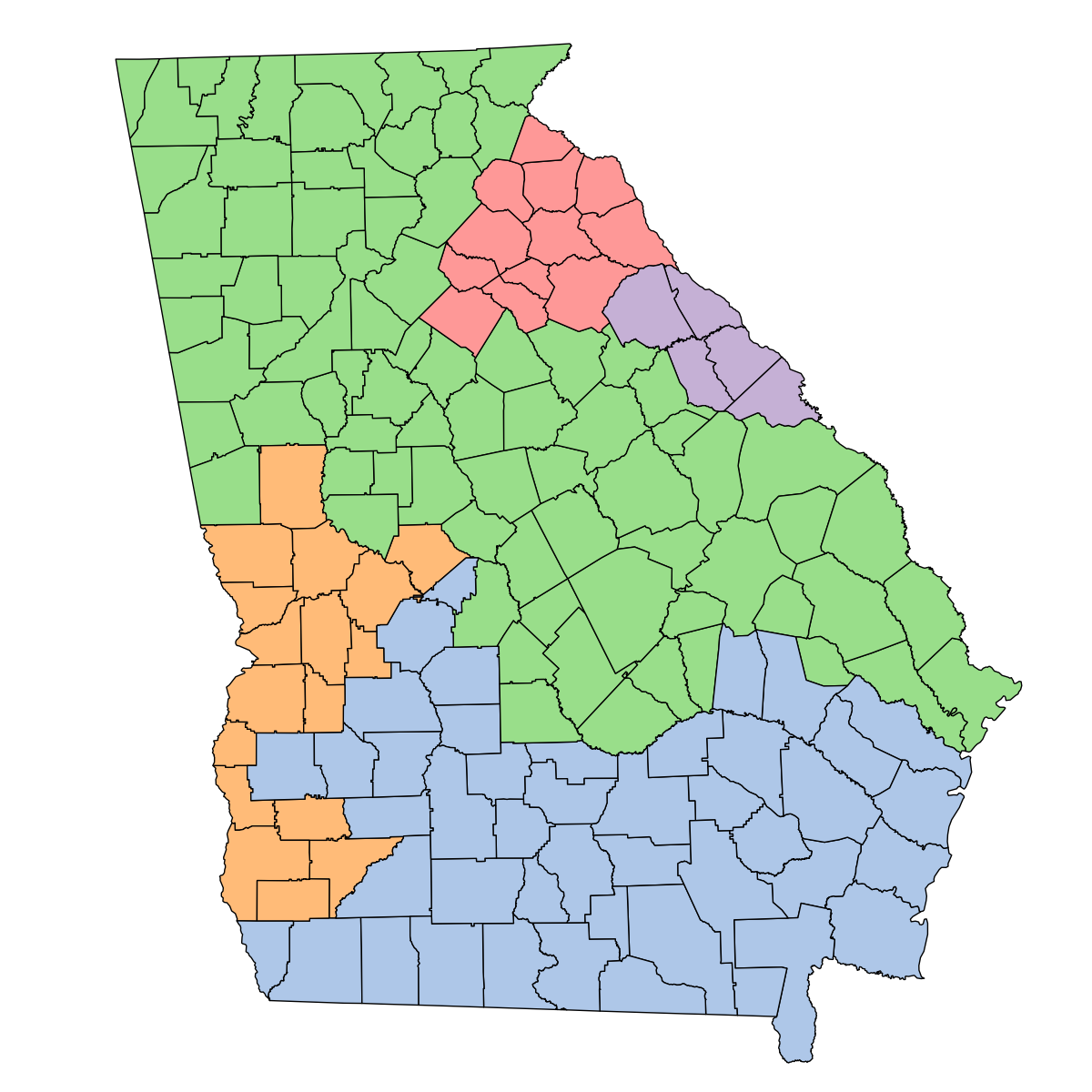}}}
\caption{Regime delineation results of the Georgia dataset. Each color represents a reconstructed region. We report the best solution among 10 repeating runs for two-stage K-Models, AZP and Regional-K-Models. (a) two-stage K-Models, SSR=25.02; (b) AZP, SSR=28.18; (c) Regional-K-Models, SSR=30.73; (d) GWR-Skater, SSR=49.29; (e) Skater-reg, SSR=40.80. SSR is calculated with standardized independent and dependent variables.}
\label{figgeog}
\end{figure}

\begin{figure}
\centering
\subfigure[Intercept]{\resizebox*{3.12cm}{!}{\includegraphics{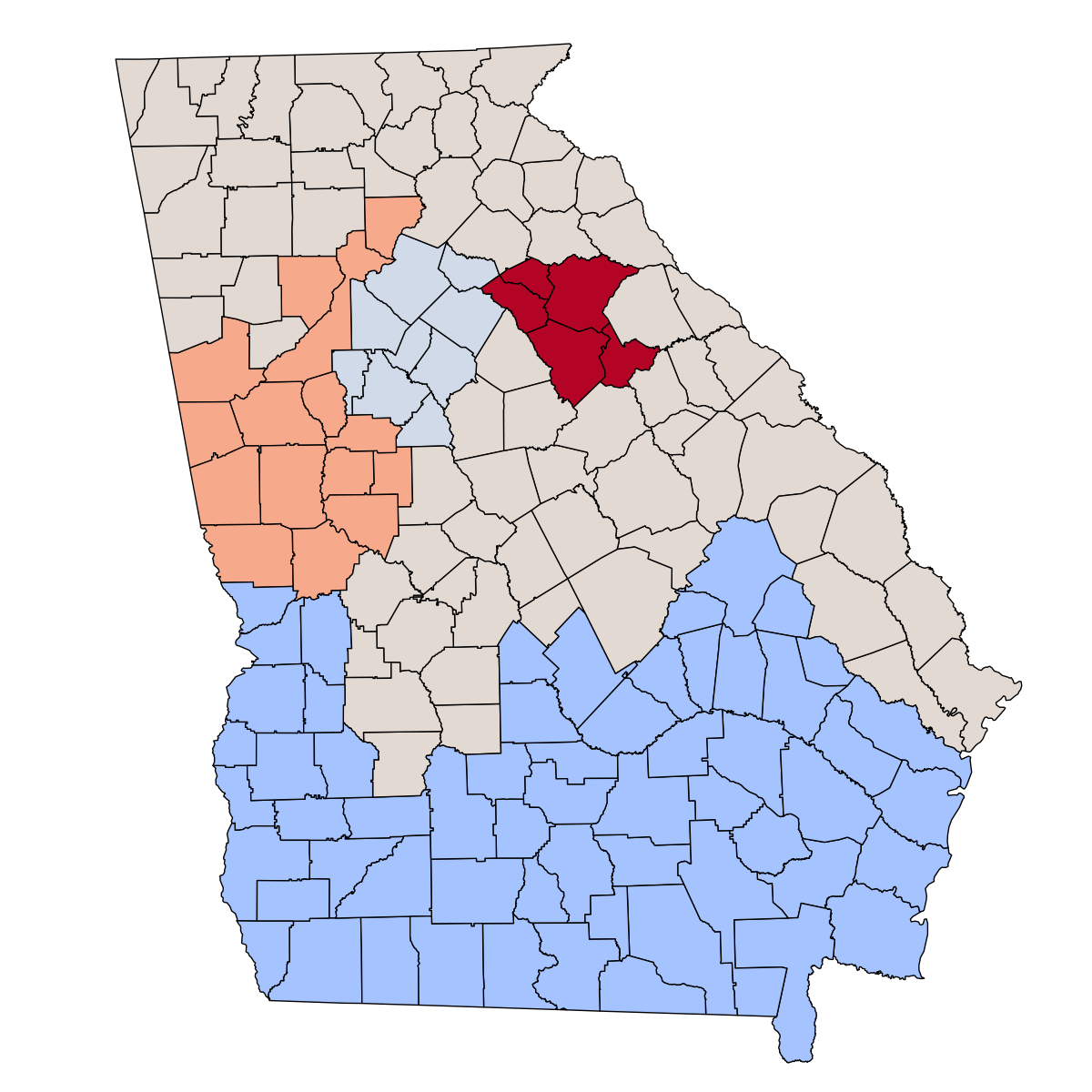}}}
\subfigure[PctFB]{\resizebox*{3.12cm}{!}{\includegraphics{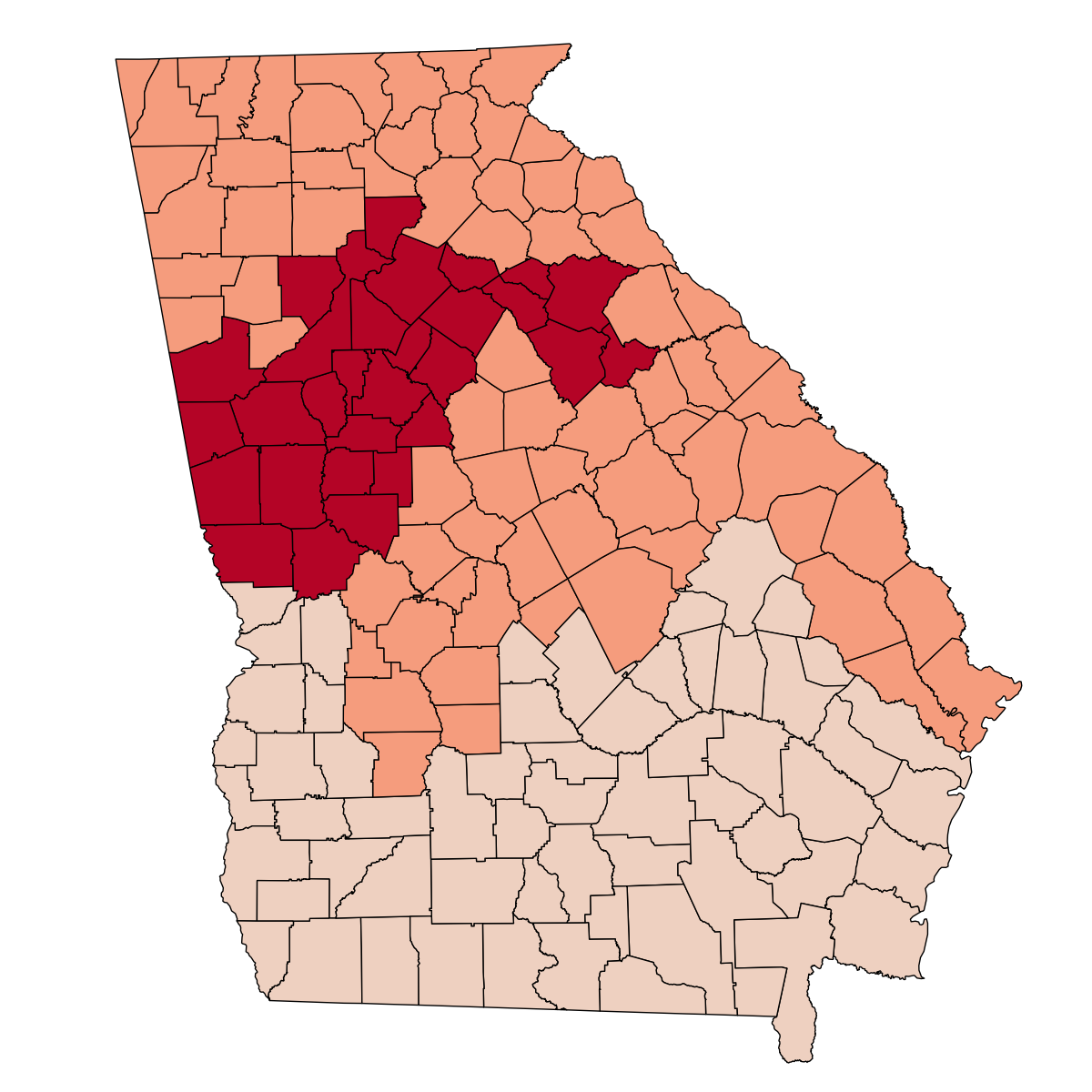}}}
\subfigure[PctBlack]{\resizebox*{3.12cm}{!}{\includegraphics{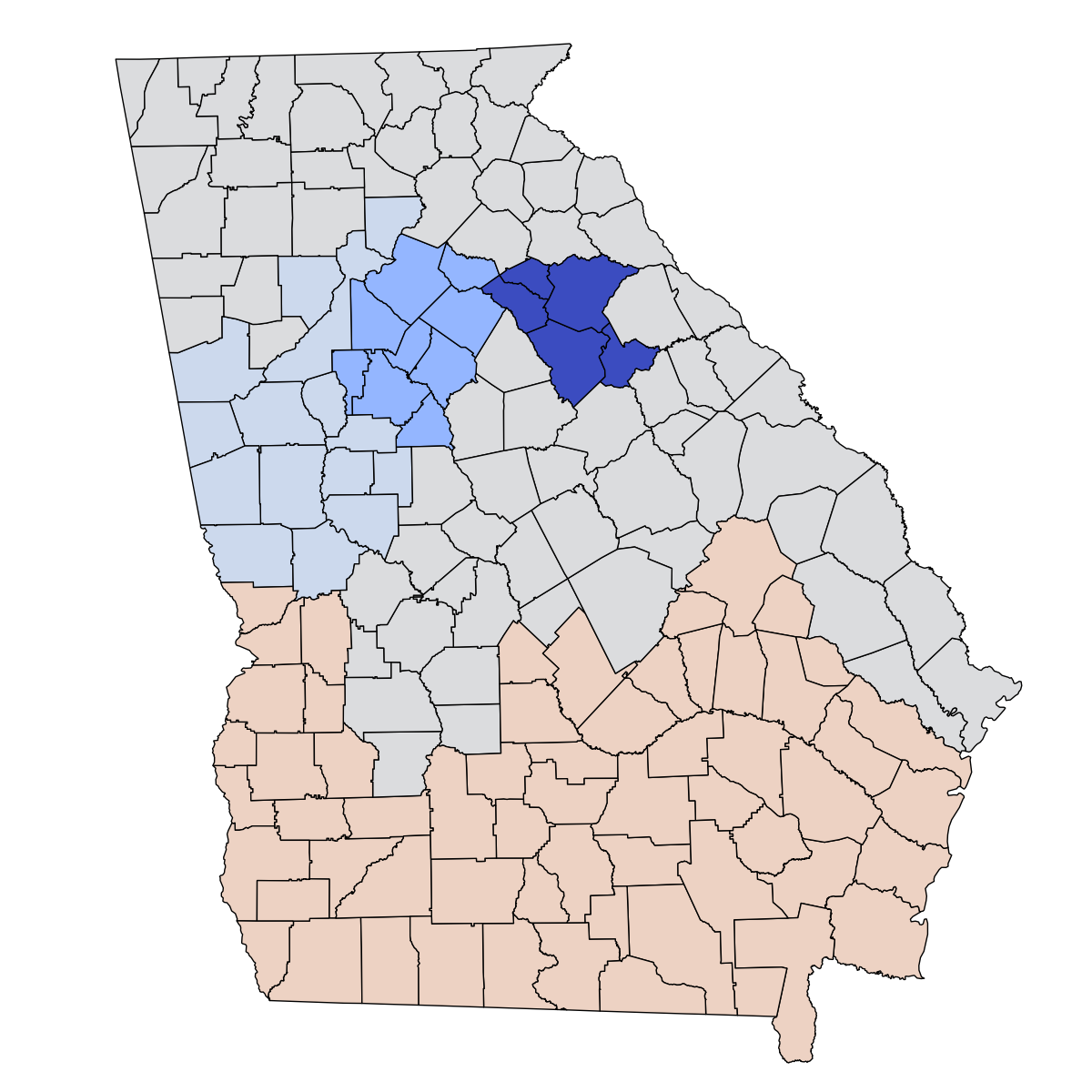}}}
\subfigure[PctRural]{\resizebox*{4.16cm}{!}{\includegraphics{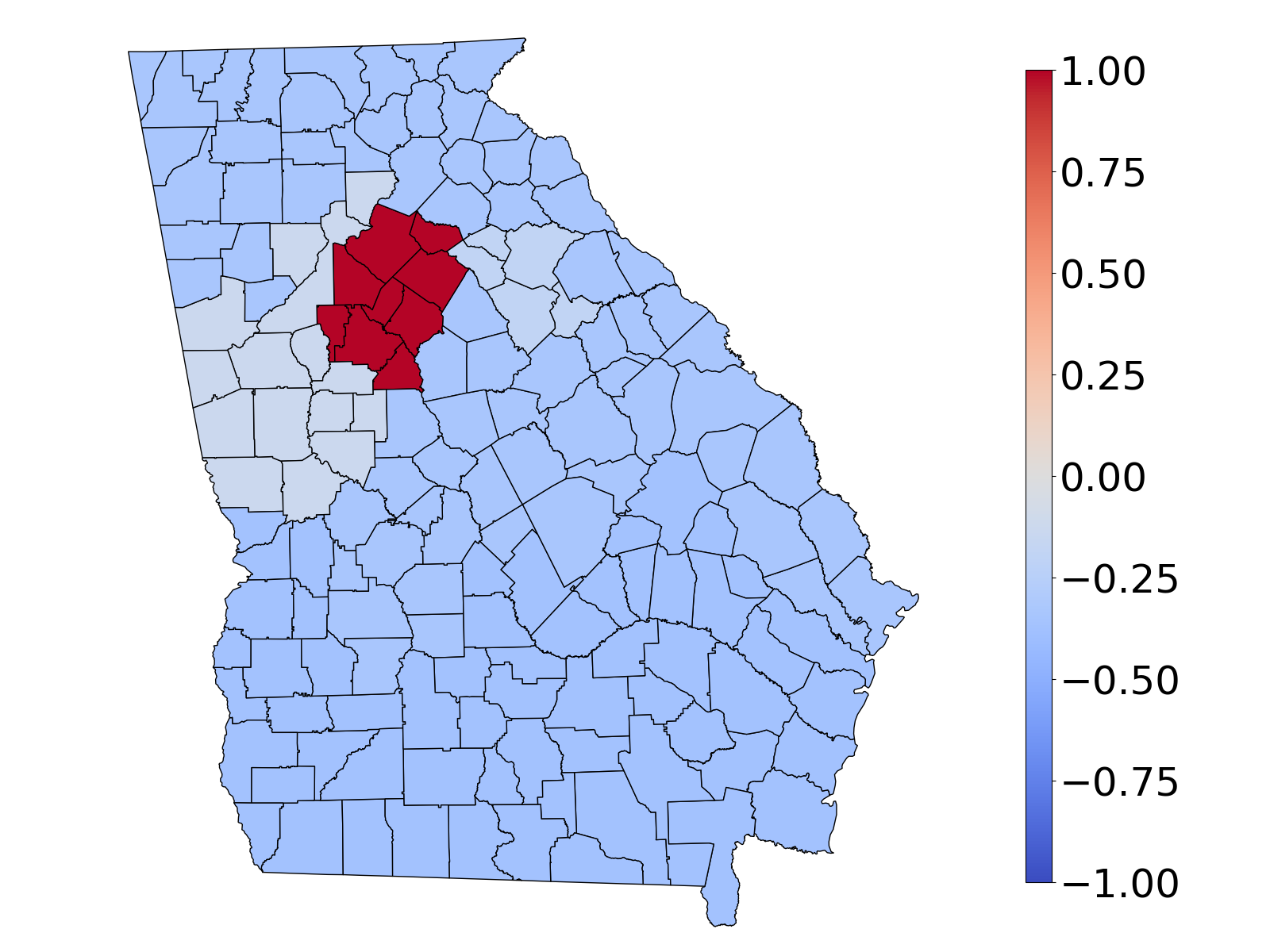}}}
\caption{Estimated regression coefficients from the K-Models algorithm. The colors indicate the magnitude and direction of the associations (blue: negative; red: positive). }

\label{figgeokm}
\end{figure}

We investigate the regime delineation at $p=5$ as an example (Figure \ref{figgeog}). All three algorithms show a general discrepancy along the north-south axis. We focus on the result of the two-stage K-Models algorithm, which performs best (lowest SSR). Figure \ref{figgeokm} shows the estimated coefficients on standardized data for each region. For all but the smallest region, an overall F-test of the regional regression model appears significant at 1\% level. Considering the overall trends, PctBach is positively associated with PctFB, and the association is stronger in the northern area. PctRural appears to be negatively associated with PctFB across Georgia. The association of PctBlack is positive in the south, and negative in the north. Two-stage K-Models also identifies two small outlier regions in the north, containing 9 and 5 counties, respectively. The extreme coefficients in both regions may indicate locally varying relationships between considered variables. The overall pattern of associations is consistent with the results from GWR \citep{YFL20}. However, GWR fails to identify the two outlier regions.

\subsection{The King County house price dataset}

House price modeling has been a major application of spatial regime regression in spatial econometrics \citep{HBHL13,BCMM16,BBP17,AnAm21}. In a hedonic house price model, house price is explained by house characteristics, including living area, number of rooms, building age, and neighborhood characteristics. Here endogenously delineated spatial regimes may correspond to housing submarkets. 

We use a house price dataset of King County, Washington provided by \cite{AnAm21}. The dataset contains 21,613 house sale records from May 2014 to May 2015. Each record contains the price and house characteristics, as well as the geographic coordinates of the house. To clean the dataset, we employ a procedure similar to that used by \cite{AnAm21}. For multiple sales at the same location, we only retain the latest record. We removed all records on the Vashon-Maury Island (which would create problems with the spatial weights), as well as several remote locations in the east which are too far from the others. Besides, anomalous records with no bedroom or bathroom are also removed. Our final dataset contains 20,616 observations. 

We apply the two-stage K-Models algorithm to delineate housing submarkets, which exhibits the best general performance and the shortest running time among the three proposed algorithms. The Skater-reg is used for comparison. We do not compare GWR-Skater because of the high computational cost to select GWR bandwidth on such a large dataset\footnote{The GWR estimation did not complete within 30 minutes on our machine.}. The dependent variable is the logarithm (base 10) of house price, and the set of 16 independent variables are the same as those used in \cite{AnAm21}. We use a k-nearest neighbor (KNN) spatial weight matrix with $k=18$, which is the minimum number of neighbors to make all observations spatially connected. The spatial weight matrix is then transformed into a symmetric matrix, which is required for generating initial regions. We set $p=5$ (as suggested by \cite{AnAm21}) and $\text{min}\_\text{obs}=20$ for both algorithms; $K=10$ for two-stage K-Models. 

\begin{figure}
\centering
\subfigure[Two-stage K-Models]{\resizebox*{6cm}{!}{\includegraphics{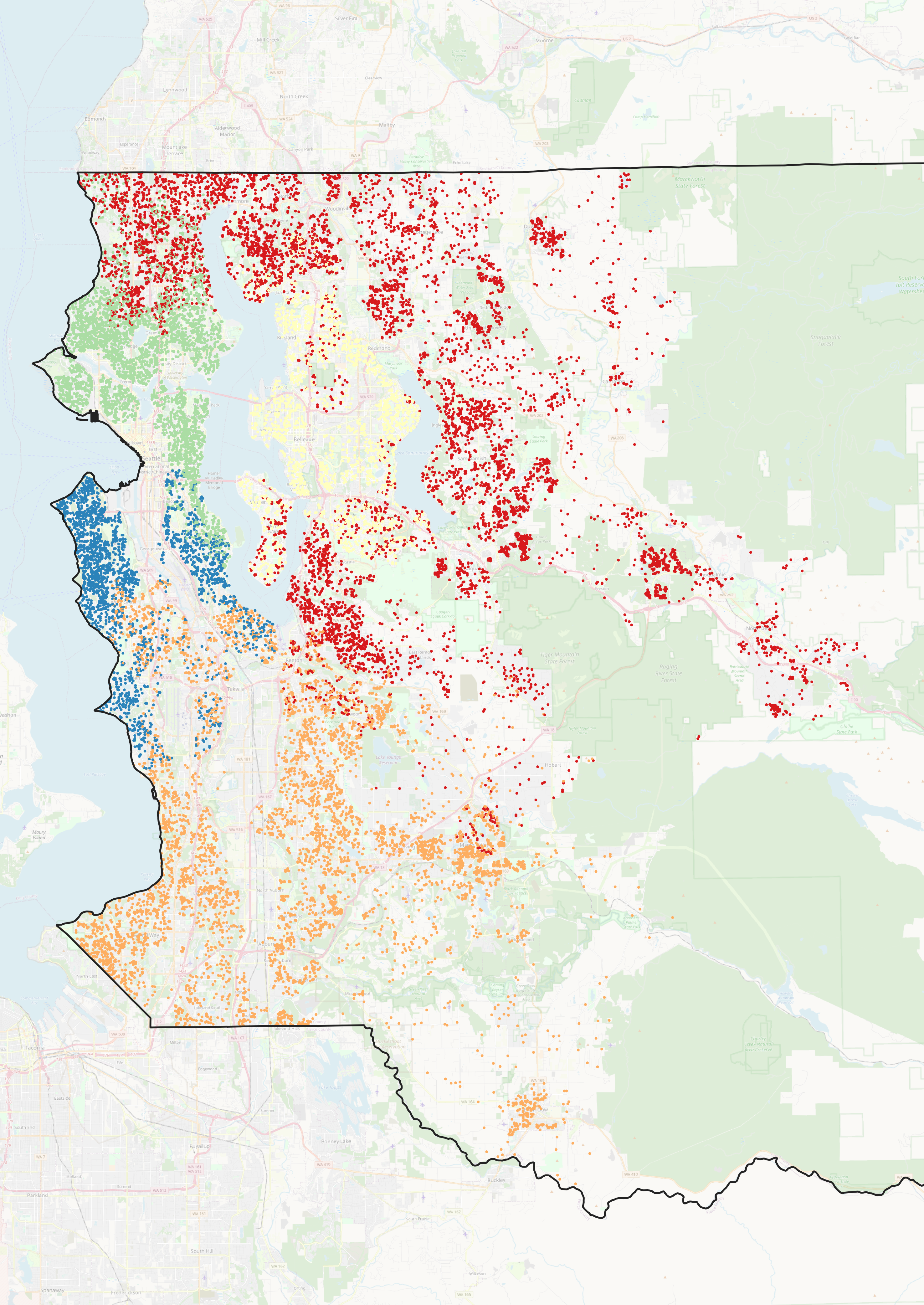}}}
\subfigure[Skater-reg]{\resizebox*{6cm}{!}{\includegraphics{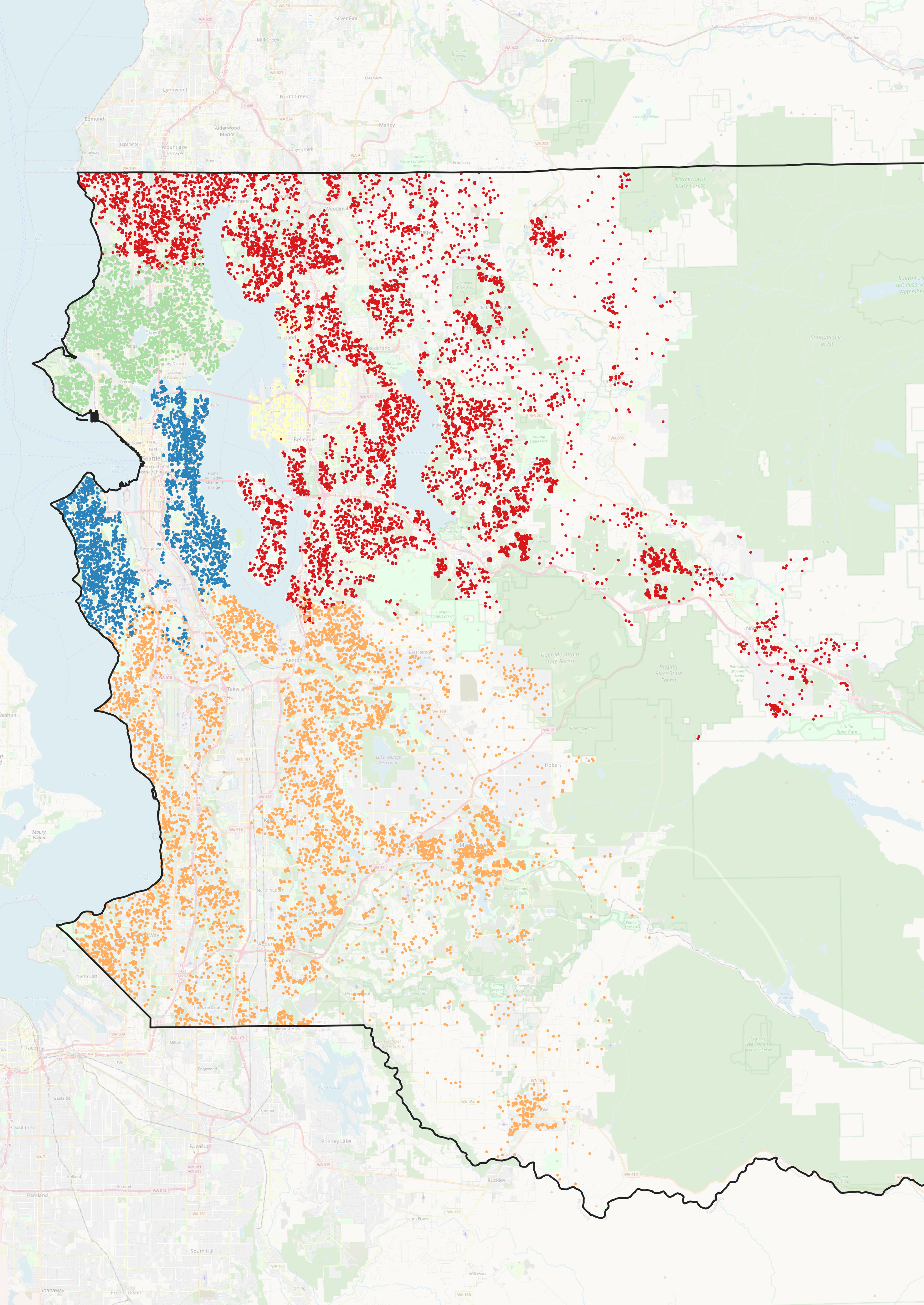}}}
\caption{Regime delineation of the King County house price dataset. Each color represents a reconstructed region. The basemap is from OpenStreetMap. (a) Two-stage K-Models, SSR=2950.58; (b) Skater-reg, SSR=3514.78. SSR is calculated with standardized independent and dependent variables.  }
\label{fig:kchouse}
\end{figure}

The regime delineations from two-stage K-Models and Skater-reg are shown in Figure \ref{fig:kchouse}. Note that regions overlap in both results because of the specified KNN neighborhood, which does not imply failure to ensure region contiguity. Skater-reg finished in about 5 minutes, while the running time of two-stage K-Models is about twice\footnote{Experiments on King County house price dataset is performed on a computer with an Intel Core i5-1135G7 CPU (2.40GHz) and 16GB of memory.}. However, two-stage K-Models delineates better spatial regimes than Skater-reg, as indicated by lower total SSR. This result brings further evidence about the ability of two-stage K-Models to delineate spatial regimes, as well as its scalability to handle large datasets.

\section{Discussion and conclusions}\label{sec:dc}

Of the three proposed algorithms, the two-stage K-Models algorithm exhibits the best general performance on both synthetic and real data, largely outperforming GWR-Skater and Skater-reg. The superior performance of two-stage K-Models on synthetic data indicates its promising capability to recover latent regions associated with spatial varying relationships, especially when the relationships change abruptly at region boundaries. Moreover, two-stage K-Models is the fastest among the three proposed algorithms, and is scalable to datasets containing up to 20,000 observations. 

The performance of AZP and Regional-K-Models is also comparable (on the synthetic datasets) or superior (on the Georgia dataset) to GWR-Skater and Skater-reg, which may reflect the limitation of considering a single spanning tree. Only a subset of possible region schemes can be produced by removing edges from a single spanning tree, which may exclude the true region scheme. On the other hand, the check of region contiguity in AZP and Regional-K-Models are not only time-consuming on large datasets, but also lead to less compact regions, as observed on the synthetic datasets. By relaxing the requirements on region connectivity, the first stage of K-Models has a wider range of improving moves during iteration, which may contribute to its better performance. Hence, the idea to ensure region contiguity throughout the zoning process might need rethinking. 

The three proposed algorithms, as well as Skater-reg, are model-agnostic. Besides multiple linear regression, other classes of statistical models, even regression trees and neural networks may be accommodated into these algorithms in a similar way. However, two issues arise for more complex regression models. First, the parameter estimation would require a large number of units in each region. After region partitioning, the number of observations in a region may be inadequate. Second, training each regional model separately may be computationally expensive. Instead of developing separate models for each region, a global model with both shared and region-specific parameters \citep{XJB21} may be a promising solution to address these issues, making the state-of-the-art GeoAI models aware of spatial heterogeneity by delineating corresponding regions.

This research investigates extensions of existing regionalization algorithms to capture spatial heterogeneous processes. This regime modeling approach is a promising way to find an equilibrium between optimizing accuracy and simplicity of geographical models. Our work responds to the stress of \textit{process} over \textit{form} in geographic information science \citep{Goo04,FoSa22}, and provides a perspective on the discussion on replicability of geographical models. When we derive a model from data in one place, would it be applicable to other places? We suppose that an application scope should be determined for each model, out of which different models should be used. The result would be multiple models, each operating within a region, which can be optimized in a top-down approach with the proposed algorithms. 

Two major limitations exist for the proposed methods. First, the class of models cannot vary across regions. Ideally, different model forms may be adopted in different regions. Second, the number of regions is required as input and fixed through the optimization process. Automatic, data-driven detection of the number of regimes is not supported. Future work may extend our methods toward solutions to these issues, which would improve their flexibility in application. 

\section*{Acknowledgements}

The acknowledgement is intentionally left blank for the peer-review process.

\section*{Data and codes availability statement}

The data and codes that support the findings of this study are available on a Github repository at https://github.com/Nithouson/regreg.

\section*{Disclosure statement}

The authors declare that they have no conflict of interest. 

\section*{Funding}

This research was supported by grants from the National Natural Science Foundation of China (41830645, 41971331, 82273731), Smart Guangzhou Spatio-temporal Information Cloud Platform Construction (GZIT2016-A5-147), and the National Key Research and Development Program of China (2021YFC2701905).

\section*{Notes on contributors}

Hao Guo is currently a Ph.D. candidate at Institute of Remote Sensing and Geographic Information Systems, Peking University. He received his B.S. in Geographic Information Science and a dual B.S. in Mathematics from Peking University in 2020. His research interests include spatial analytics, geo-spatial artificial intelligence, and spatial optimization.

Andre Python is ZJU100 Young Professor in Statistics at the Center for Data Science, Zhejiang University, P.R. China. He received his B.S. and M.S. from the University of Fribourg, Switzerland and his Ph.D. from the University of St Andrews, United Kingdom. He develops and applies spatial models and interpretable machine learning algorithms to better understand the mechanisms behind the observed patterns of spatial phenomena.

Yu Liu is currently the Boya Professor of GIScience at the Institute of Remote Sensing and Geographic Information Systems, Peking University. He received his B.S., M.S., and Ph.D. degrees from Peking University in 1994, 1997, and 2003, respectively. His research interests mainly focus on humanities and social sciences based on big geo-data.

\appendix

\section{Supplementary experiments on synthetic data}

\subsection{Effect of random noise}

The standard error of the Gaussian noise $\sigma=0.1$ (low noise) is used in Section \ref{sec:synthetic}. To evaluate the effect of random noise on algorithm performance, we repeat the regime optimization experiments with $\sigma=0.2$ (medium noise) and $\sigma=0.3$ (high noise). For each simulation, the true region scheme and coefficients are retained, while new $\textbf{x},y$ arrays are generated  with increased noise levels. Results are summarized in Table \ref{tabnoi}. We only report SSR and Rand index for simplicity. The relative performance ranking of the five algorithms is similar with the condition of $\sigma=0.1$. With $\sigma =0.3$, the average RI for two-stage K-Models is still over 0.9, indicating its robustness to random noise.

\begin{table}
\tbl{Average SSR and RI over 50 simulations with different noise levels.}
{\begin{tabular*}{\hsize}{@{}@{\extracolsep{\fill}}llcccc@{}}
\toprule
\multirow{2}{*}{Dataset}& \multirow{2}{*}{Algorithm} & \multicolumn{2}{c}{$\sigma = 0.2$} & \multicolumn{2}{c}{$\sigma = 0.3$} \\ \cmidrule{3-6}
& & SSR & RI & SSR & RI\\ \midrule
\multirow{5}{*}{\textit{Rectangular}}& K-Models &	\textbf{45.00} 	&	\textbf{0.9573} 	&	\textbf{84.00} 	&	\textbf{0.9349} \\
& AZP &	206.26 	&	0.8196 	&	271.27 	&	0.8011 \\
& Reg-K-Models & 409.96 	&	0.7850 	&	429.34 	&	0.7764 \\
& GWR-Skater &	266.90 	&	0.8532 	&	289.08 	&	0.8435 \\
& Skater-reg & 417.81 	&	0.6738 	&	453.89 	&	0.6703 \\
\midrule
\multirow{5}{*}{\textit{Voronoi}}& K-Models &	\textbf{44.98} 	&	\textbf{0.9609} 	&	\textbf{82.44}	&	\textbf{0.9500} \\
& AZP &	162.54 	&	0.8360 	&	190.82 	&	0.8267 \\
& Reg-K-Models & 294.52 	&	0.8082 	&	317.06 	&	0.8024 \\
& GWR-Skater &	223.57 	&	0.8826 	&	259.18 	&	0.8740 \\
& Skater-reg & 292.25 	&	0.7662 	&	327.13 	&	0.7379 \\
\midrule
\multirow{5}{*}{\textit{Arbitrary}}& K-Models &	\textbf{66.93} 	&	\textbf{0.9334} 	&	\textbf{105.53}	&	\textbf{0.9165}\\
& AZP &	185.93 	&	0.8432 	&	231.80 	&	0.8299 \\
& Reg-K-Models & 302.98 	&	0.8255 	&	366.67 	&	0.8072 \\
& GWR-Skater &	287.50 	&	0.8425 	&	330.19 	&	0.8352 \\
& Skater-reg & 360.64 	&	0.7327 	&	379.49 	&	0.7470 \\
\bottomrule
\end{tabular*}}
\label{tabnoi}
\footnotesize
Note: K-Models is short for two-stage K-Models; Reg-K-Models is short for Regional-K-Models. The best metric values in each simulation setting is put in bold.
\end{table}

\subsection{Algorithm Stability}

Unlike GWR-Skater and Skater-reg, the three proposed algorithms include some degree of randomness in the optimization process. Randomness is induced by random region growth in region initialization, random choice among candidate moves, and the order in which algorithms execute checks throughout all regions or units. In practice, repeated runs could be performed, of which the best solution is retained. 

\begin{figure}
\centering
\subfigure[]{\resizebox*{7cm}{!}{\includegraphics{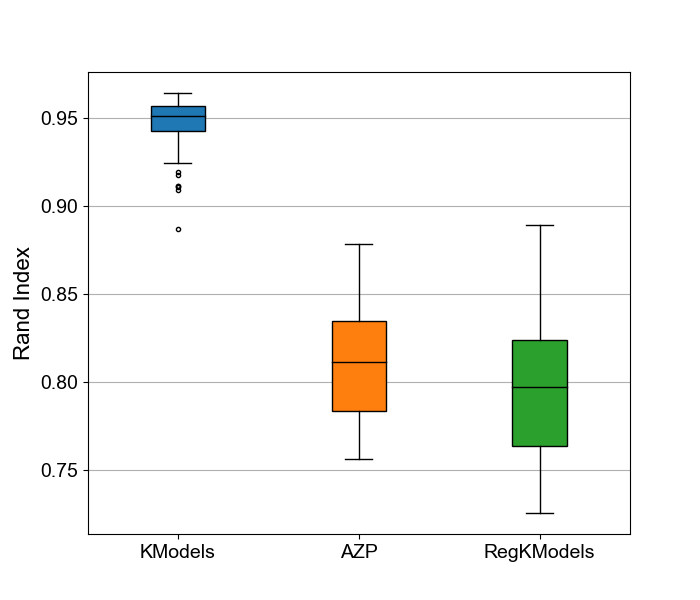}}}
\subfigure[]{\resizebox*{7cm}{!}{\includegraphics{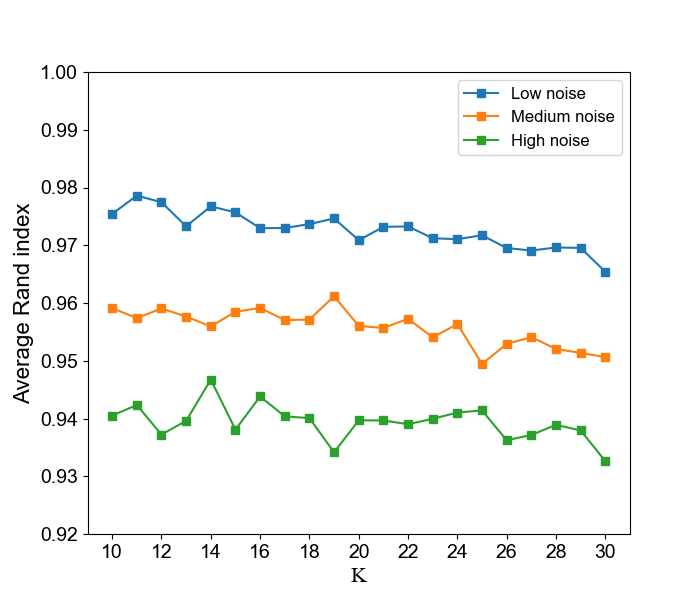}}}
\caption{Results on algorithm stability and effect of parameter $K$ in two-stage K-Models. (a) Boxplot showing Rand index values of solutions produced by three algorithms (K-Models (blue), AZP (orange), Regional-K-Models (green)). Each algorithm is executed 50 times on the  same data. (b) Average Rand index of solutions produced by two-stage K-Models at different $K$ values. The \textit{Rectangular} dataset with three noise levels are used, including low noise (blue, $\sigma = 0.1$), medium noise  (orange, $\sigma = 0.2$), high noise (green, $\sigma = 0.3$). The algorithm runs on 50 simulations for each noise level.} 
\label{figsupp}
\end{figure}

To examine the stability of the proposed algorithms, we take one simulation from the \textit{Rectangular} dataset, and repeat the regime optimization experiment with each algorithm for 50 times. Figure \ref{figsupp}a shows the distributions of Rand index values. Two-stage K-Models shows less variability in repeated runs, indicating better stability compared to AZP and Regional-K-Models. Even considering 50 repeats, few solutions from AZP and Regional-K-Models are comparable with two-stage K-Models. 

\subsection{Effect of $K$ in two-stage K-Models}

We investigate selection of the parameter $K$ in two-stage K-Models. We use the \textit{Rectangular} dataset with three noise levels ($\sigma=0.1,0.2,0.3$, respectively). Figure \ref{figsupp}b shows the average Rand index over 50 simulations at different $K$ values in $[10,30]$. Results show that the algorithm performance is not sensitive to the choice of $K$, although the Rand index slightly decreases if a very large $K$ value is used. 

\end{document}